\documentclass[showpacs,superscriptaddress,preprintnumbers,amsmath,
amssymb,nofootinbib]{revtex4}
\usepackage{amsmath,amssymb,latexsym}
\usepackage{graphicx,bm}
\usepackage{dcolumn}    
\usepackage{epsf,color}

\definecolor{mod1}{rgb}{1,0,0}
\definecolor{mod2}{rgb}{0,0,1}
\definecolor{mod3}{rgb}{0,.5,0}
\definecolor{modAT}{rgb}{0.5,0,0.5}

\begin{document}
%
%
%
%
%
%
\title{ 
{Detecting a stochastic background of gravitational waves \\
     in the 
 presence of non-Gaussian noise\\
 {\it A performance of generalized cross-correlation statistic}} }
%
%
%
%
%
%
\author{Yoshiaki Himemoto} \email{himemoto_at_utap.phys.s.u-tokyo.ac.jp}
\affiliation{ Department of Physics, The University of Tokyo, Tokyo
  113-0033, Japan }

\author{Atsushi Taruya} \email{ataruya_at_phys.s.u-tokyo.ac.jp}
\affiliation{ Research Center for the Early Universe~(RESCEU), School
  of Science, The University of Tokyo, Tokyo 113-0033, Japan }

\author{Hideaki Kudoh} \email{kudoh_at_utap.phys.s.u-tokyo.ac.jp}
\affiliation{ Department of Physics, The University of Tokyo, Tokyo
  113-0033, Japan }
\affiliation{ Department of Physics, University of California, Santa Barbara, CA 93106, USA}

\author{Takashi Hiramatsu} \email{hiramatsu_at_utap.phys.s.u-tokyo.ac.jp}
\affiliation{ Department of Physics, The University of Tokyo, Tokyo
  113-0033, Japan }

\preprint{UTAP-558, RESCEU-16/06}

%
%
%
%
%
%
%
%
\begin{abstract}

We discuss a robust data analysis method to detect 
a stochastic background of gravitational waves in the presence of 
non-Gaussian noise. In contrast to the standard cross-correlation (SCC) 
statistic frequently used in the stochastic background searches, 
we consider a {\it generalized cross-correlation} (GCC) statistic, 
which is nearly optimal even in the presence of non-Gaussian noise. 
The detection efficiency of the GCC statistic is investigated 
analytically, particularly focusing on the statistical 
relation between the false-alarm and the false-dismissal probabilities, 
and the minimum detectable amplitude of gravitational-wave signals.
We derive simple analytic formulas for these statistical quantities. 
The robustness of the GCC statistic is clarified based on these formulas, 
and one finds that the detection efficiency of the GCC statistic roughly 
corresponds to the one of the SCC statistic neglecting the contribution 
of non-Gaussian tails. This remarkable property is checked by performing 
the Monte Carlo simulations and successful agreement between analytic 
and simulation results was found.

\end{abstract}

\maketitle

\section{Introduction}
\label{sec:intro}

A stochastic background of gravitational waves is expected to be 
very weak among various types of gravitational-wave signals. 
Such a tiny signal is produced by an incoherent superposition of 
many gravitational-wave signals coming from the irresolvable
astrophysical objects and/or diffuse 
high-energy sources in the early universe. 
Up to now, various 
mechanisms to produce stochastic signals have been proposed and 
their amplitudes and spectra are estimated quantitatively 
(for the review see Ref.~\cite{allen96,Maggiore:1999vm}).

Despite the small amplitude of the signals, the stochastic backgrounds of 
gravitational waves 
contain valuable cosmological information about cosmic expansion 
history and astrophysical phenomena. Because of its weak interaction, the extremely early stage of the universe beyond the last 
scattering surface of the electromagnetic waves would be probed 
via the direct detection of inflationary gravitational-waves background. 
In this sense, gravitational-wave backgrounds are an ultimate 
cosmological tool and the direct detection of such signals will open a 
new subject of cosmology.

As a trade-off, detection of stochastic background is very difficult and the
challenging problem. Recently, the observational bound of stochastic 
background has been updated by 
Laser Interferometer Gravitational Wave Observatory 
(LIGO \cite{ligo}) third scientific run \cite{Abbott:2005ez} and the 
amplitude of signal is constrained to 
$\Omega_{\rm gw} \lesssim
8.4 \times 10^{-4}$, where $\Omega_{\rm gw}$ is the energy density of 
gravitational wave divided by the critical energy density. 
While this is the most stringent constraint obtained from the laser 
interferometer \cite{Abbott:2004rt}, this bound is still larger than the limit inferred from 
the big-bang nucleosynthesis. Hence, for the direct detection of 
stochastic signals, a further development to increase 
the sensitivity is essential. To do this, one obvious approach 
is to construct a more sophisticated detector whose sensitivity level 
is only limited by the quantum noises. Next-generation of ground-based 
detectors, such as LIGO II and Large-scale Cryogenic Gravitational-wave 
Telescope (LCGT) \cite{Mio:2003ii}, will greatly improve the sensitivity that 
reaches or may beat the standard quantum limit. 
Furthermore, the space-based interferometer will be suited to prove gravitational wave backgrounds due to its lower observational band~\cite{Kudoh:2005as}.
Another important direction is to explore the efficient and the robust technique of data analysis for signal detection.

In this paper, we shall treat the latter issue, particularly focusing 
on  the signal detection in the presence of the non-Gaussian noises. 
When we search for the weak stochastic signals embedded in the 
detector noise, we have no practical way to discriminate between 
the detector noise and a stochastic signal by using only a single 
detector. To detect a stochastic signal, 
we must combine the two outputs at different detectors 
and quantify the statistical correlation between them.  
This cross-correlation technique is the robust statistical method that 
is still useful in the cases with large detector noises. 
The so-called standard cross correlation technique
has been frequently used in the data analysis of laser interferometers. 
Note that the standard cross correlation statistic 
was derived under the assumption that both the signals and the 
instrumental noises obey stationary Gaussian process 
\cite{Christensen:1992wi,Flanagan:1993ix,Allen:1997ad}. 
In practice, however, gravitational wave 
detectors do not have a pure Gaussian noise. 
Because of some uncontrolled mechanisms,  
most experiments exhibit a non-Gaussian tail. In the presence of 
non-Gaussianity, the direct application of the standard cross-correlation 
statistic significantly degrades the sensitivity of 
signal detection. A more appropriate cross-correlation 
statistic to reduce the influence of the non-Gaussian tails 
should be desirable in the data analysis of signal detection.

In Refs.~\cite{Allen:2001ay,Allen:2002jw}, the standard 
cross-correlation analysis was extended to deal with more realistic
situation. They found that such a modified statistic shows a 
better performance compared to 
the standard cross-correlation statistic \cite{Allen:2001ay}. 
This modified statistic is called the locally optimal 
statistic \cite{Saleem}. Roughly speaking, 
the usual standard cross-correlation statistic uses all detector samples, 
while the locally optimal statistic excludes the samples of the non-Gaussian 
tails outside the main Gaussian part from the detector samples. As 
a result, 
the statistical noise variance in the locally optimal statistic becomes 
small due to the truncation of the samples of the non-Gaussian tail,  
so that the effective signal-to-noise ratio becomes large.

In this paper, we derive analytical formulas for the 
false-alarm and the false-dismissal probabilities and the minimum
detectable signal amplitude to quantify the performance 
of the locally optimal statistic. 
Then, we demonstrate the detection efficiency of locally optimal statistic 
in a simple non-Gaussian noise model, in which the probability distribution 
of the instrumental noise is described by 
the two-component Gaussian noise. 
Based on the analytical formulas, 
the efficiency of the locally optimal statistic is quantified 
compared to the standard cross-correlation statistic.

The structure of this paper is as follows. In the next section, we 
briefly review the detection strategy for a stochastic background. 
We then introduce the generalized cross-correlation statistic which is
nearly optimal in the presence of non-Gaussian noise.
In Sec.\ref{sec:3}, particularly focusing on 
the two-component Gaussian model  as 
a simple model of non-Gaussian noises, 
we analytically estimate the false-alarm and 
the false-dismissal probabilities. Based on this, we 
obtain the analytic expression for the minimum detectable 
amplitude of stochastic signals. 
The resultant analytic formulas imply that
the detection efficiency of the GCC statistic 
roughly corresponds to the one of the SCC statistics neglecting the 
contribution of non-Gaussian tails.
These remarkable properties are checked and confirmed by 
performing the Monte Carlo simulations in Sec.\ref{sec:4}. 
Finally, in Sec.\ref{sec:5}, we close the paper with a summary 
of results and a discussion of future prospects.

\section{Optimal Detection Statistic in the 
presence of non-Gaussian Noise}
\label{sec:2}

As we previously mentioned, 
the gravitational-wave background (GWB) signal 
is expected to be very week and is usually masked 
by the detector noises. 
To detect such tiny signals, it is practically impossible to 
detect the GWB signal from the single-detector measurement. 
Thus, we cross-correlate the two outputs obtained from 
the different detectors and seek a common signal.  
We denote the detector outputs by $s_{i}^{k}$ with
\begin{equation}
s_{i}^{k}=h_{i}^{k}+n_{i}^{k}, \quad (i=1,2,~~~k=1,\cdots,N),  
\label{output}
\end{equation}
where $i=1,2$ labels the two detectors, and $k=1,\cdots,N$ is a time index.
Here, $h_{i}^{k}$ is the gravitational-wave signal, whose amplitude is 
typically $\epsilon$, and $n_{i}^{k}$ is the noise 
in each detector. The $N \times 2$ output matrix ${\cal S}$ is made up 
of these outputs. Throughout this paper, 
we discuss the optimal detection method 
under the assumption of weak signal, i.e., 
$|h_i^k| \sim \epsilon \ll |n_i^k|$.

\subsection{Detection statistic}

To judge whether a gravitational signal is indeed present in detector
outputs or not, 
the simplest approach is to use a detection statistic 
$\Lambda = \Lambda({\cal S})$. 
When $ \Lambda $ exceeds a threshold $\Lambda^{\ast}$, we think that the 
 signal is detected, and not detected otherwise. 
The statistic $\Lambda$, which is made up of random variables ${\cal S}$, 
exhibits random nature under the finite sampling and because of this, 
we have two types of error depending on the detection 
criterion $\Lambda^{\ast}$. The probabilities of these errors  
are often called {\it false-alarm rate} and {\it false-dismissal rate}. 
The probability of the false alarm is the one that 
we conclude to have detected a signal, but the signal is in fact absent.  
We denote the probability by $P_{\rm FA}[\Lambda^{\ast}]$. On the other hand, 
the probability of the false dismissal which we denote by 
$P_{\rm FD}[\Lambda^{\ast}]$ is the probability that we fail to 
detect a signal even though the signal is in fact present. 
Thus, one may say that the detection statistic 
is {\it optimal} only when the two errors are minimized.  
Neyman and Pearson showed that the likelihood
ratio is the optimal decision statistic that minimizes $P_{\rm FD}$ 
for a given value of 
$P_{\rm FA}$ \cite{Neyman_Pearson}. The likelihood ratio is given by 
\begin{equation}
\Lambda = \frac{p({\cal S}|\epsilon)}{p({\cal S}|0)}. 
\label{likelihood}
\end{equation}
Here, the quantity $p({\cal S}|\epsilon)$ is the probability distribution 
function of the observational data set ${\cal S}$ in 
the presence of the signal, whose amplitude is given by $\epsilon$. 
We are specifically concerned with the detection of weak signals.
In such a situation, regarding $\epsilon$ as a small parameter,
one can expand $\Lambda$ as
\begin{equation}
\Lambda = 1+\epsilon \Lambda_{1}+ \epsilon^{2} \Lambda_{2}+O(\epsilon^{3}).
\label{expansion}
\end{equation}
As long as $\epsilon$ is small, the higher-order terms of
$O(\epsilon^{2})$ are neglected and 
the quantity $\Lambda_{1}$ approximately becomes the
optimal decision statistic. This statistic is called the {\it locally 
optimal statistic} \cite{Allen:2001ay}. If $\Lambda_{1}$ becomes zero, 
then $\Lambda_{2}$ is the optimal decision statistic.

\subsection{Standard and generalized cross-correlation statistics}
\label{subsec:SCC_GCC}

In order to obtain some insights into the locally 
optimal statistic, we consider the simplest situation
for the data analysis of signal detection.  
For any two detectors, we assume that their orientations
are coincident and
coaligned without any systematic noise correlation between them, 
so that two detectors receive the same signal, i.e., 
$h_{1}^{k}=h_{2}^{k} = h^{k}$. 
There are several missions that realize such a situation.
The ongoing LIGO project has two colocated detectors in the
Hanford site, although the arm length of each detector
is different \cite{ligo}. 
The LCGT detector proposed by 
the Japanese group also has two colocated detector sharing a common arm
cavity \cite{Mio:2003ii}.

In addition to the orientation of the detectors, 
we further assume that each detector has a white and 
stationary noise. In this case, the joint probability distribution of the 
detector noises is given by 
\begin{equation}
 p_{n}({\cal N})=\prod_{k=1}^{N}e^{-f_{1}(s_{1}^{k}-h^{k})
-f_{2}(s_{2}^{k}-h^{k})},
\label{npdf0}
\end{equation}
where the symbol ${\cal N}$ represents 
the noise contribution to the output matrix 
${\cal S}$. Note that 
Eq.(\ref{npdf0}) reduces to a multivariate Gaussian distribution 
if the function $f_{i} $ 
becomes quadratic in its argument.  Thus, 
the function $f_{i} $ other than 
the quadratic form implies the non-Gaussianity of the detector noises. 
As for the probability of the signal amplitude, we also assume 
that the signal is white,
so that the probability distribution function for ${\cal H}={h^{1},....,h^{N}}$ is
expressed by
\begin{equation}
 p_{h}({\cal H})
=\prod_{k=1}^{N} p_{h^{k}}(h^{k}).
\label{spdf0}
\end{equation}
From Eqs.~(\ref{output}), ({\ref{npdf0}}) and ({\ref{spdf0}}),
 the numerator in the likelihood ratio (\ref{likelihood})
is given by
\begin{equation}
p({\cal S}|\epsilon)=\int dh^{1} \, \cdots \, \int dh^{N}
\, p_{h}({\cal H}) \,  p_{n}({\cal N}).
\end{equation}
Expanding the likelihood ratio with respect to 
$|h^{k}|\sim\epsilon \ll 1$ around 
zero, we obtain the locally optimal
statistic \cite{Allen:2001ay,Allen:2002jw}. 
In the present case, $\Lambda_{1}$ in Eq.~(\ref{expansion}), 
which includes the linear term of the signal,
vanishes because the stochastic
gravitational wave is usually a zero-mean signal. 
Therefore, $\Lambda_{2}$ turns out to be the 
optimal {\it decision statistic}. $\Lambda_{2}$ is composed 
of second derivative terms 
and some quadratic of the first derivative
terms with respect to $s_{i}^{k}$. We then classify these terms
into {\it single-detector} statistic and {\it two-detector}
statistic \cite{Allen:2001ay}. The former statistic, which is described
by quantities such as $f_{i}^{''}$ and $(f_{i}^{'})^{2}$,
are only relevant in the cases when 
the gravitational-wave signal dominates the detector noises. 
The latter two-detector statistic is given by \cite{Allen:2001ay}
\begin{equation}
 \Lambda_{\rm GCC} \propto \frac{1}{N} \sum_{k=1}^{N} f_{1}'(s_{1}^{k}) \,
 f_{2}'(s_{2}^{k}),
\label{gcc}
\end{equation}
where we used the fact that the signal is white. 
In this paper, we especially call it 
generalized cross-correlation (GCC) statistic\footnote{Although 
we extracted the cross-correlation term 
by hand, the Bayesian derivation
automatically eliminates the self-correlation terms \cite{Allen:2002jw}.}.

In what follows, for the purpose of our analytic study,
we treat the non-Gaussian parameters in the function $f_{i}$ as known
parameter. Furthermore, we define the counterpart of the GCC statistic
in the absence of non-Gaussianity 
as:
\begin{equation}
\Lambda_{\rm SCC} \equiv 
\frac{1}{N} \sum_{k=1}^{N} s_{1}^{k} s_{2}^{k}\,,
\label{scc}
\end{equation}
which we call the standard cross-correlation (SCC) statistic. 
Strictly speaking, the decision statistics derived here are 
not the optimal decision statistics \cite{Allen:1997ad, Drasco:2002yd}. 
For instance, in the case of the Gaussian noises with 
unknown variances, the optimal decision statistic differs from 
Eq. (\ref{scc}) by a factor $(\hat{\sigma}_1\hat{\sigma}_2)^{-1}$, where 
$\hat{\sigma}_i$ is the square-root of 
the autocorrelation function for the output 
signals $s_i^k$, i.e., $\hat{\sigma}_i^2\equiv(1/N)\sum_k (s_i^k)^2$. 
Nevertheless, in the large-sample limit $N\to\infty$, 
statistical fluctuations in the autocorrelation function 
become negligible relative to those in $\Lambda_{\rm SCC}$ 
and the autocorrelation functions can be treated as 
constants. Thus, in the limit $N\to\infty$, the factor 
$(\hat{\sigma}_1\hat{\sigma}_2)^{-1}$ is irrelevant and one can 
identify Eq. (\ref{scc}) as the optimal decision statistic 
\cite{Drasco:2002yd}. In this sense, 
Eq. (\ref{scc}) may be regarded as an {\it nearly} optimal statistic. 
Although it seems difficult to prove that the statistic $\Lambda_{\rm GCC}$ 
really approaches the (locally) optimal statistic in the large-sample limit 
with the non-Gaussian noises, the essential properties in the statistic 
$\Lambda_{\rm GCC}$ is the same as those in the optimal decision 
statistic derived from Bayesian treatment \cite{Allen:1997ad}.  
We hope that the resultant analytic formulas for detection efficiency 
are also useful in the practical situation 
that we do not know the noise parameters {\it a priori}.

\section{Analytic estimation of the  detection efficiency}
\label{sec:3}

\begin{figure}[!b]
\begin{center}
\includegraphics[height=6.5cm,angle=0,clip]{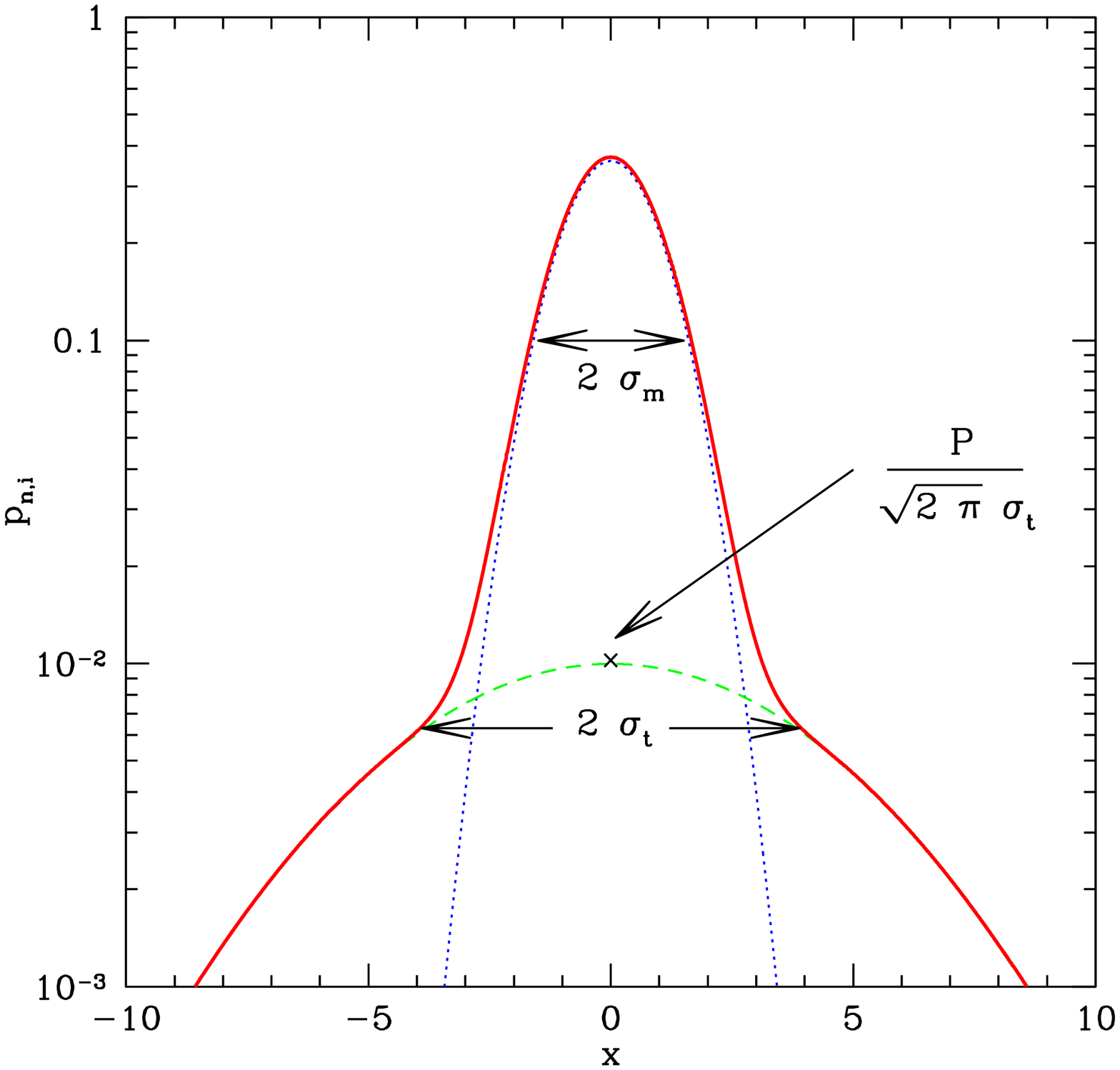}
\hspace*{0.5cm}
\includegraphics[height=6.5cm,angle=0,clip]{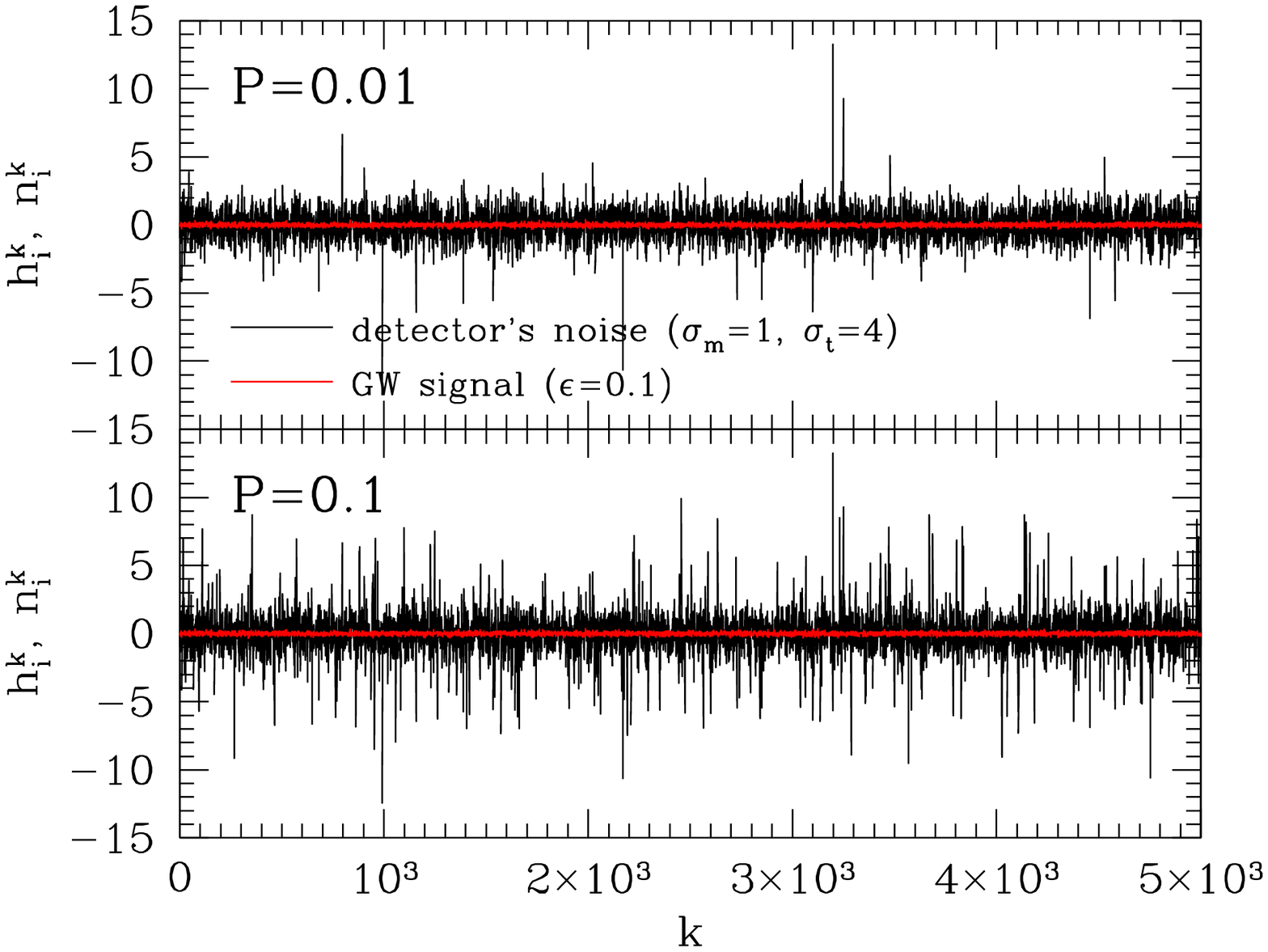}
\end{center}
    \caption{{\it Left}: Probability distribution function of instrumental 
      noise given by Eq.(\ref{npdf1}). The model parameters are 
      set to $\sigma_{\rm m}=1$, $\sigma_{\rm t}=4$ 
and $P=0.1$. Here we dropped
 the detector label $i$. 
      {\it Right}: 
      Time-series data of the non-Gaussian noise generated by 
      Eq. (\ref{npdf1}). Top panel  
      shows the result with tail fraction $P=0.01$, while the bottom 
      panel plots the case with larger value, $P=0.1$. 
      In both panels, we 
      specifically set the model parameters $\sigma_{\rm m}$ and 
      $\sigma_{\rm t}$ as $\sigma_{\rm m}=1$ and $\sigma_{\rm t}=4$. 
      For comparison, we also plot the weak signal of the stochastic 
      gravitational waves with $\epsilon=0.1$. }
 \label{fig:time_series}
\end{figure}

We wish to clarify how the GCC statistic improves the detection
efficiency in the presence of non-Gaussian noise in an analytic way.   
For this purpose, we treat the simple non-Gaussian model, in which 
the probability distribution of the detector noises is characterized 
by the two-component Gaussian distribution 
given by  \cite{Allen:2001ay,Creighton:1999qw}:
\begin{eqnarray}
 p_{n,i}(x)&=&
e^{-f_{i}(x)}=\frac{(1-P_{i})}{\sqrt{2\pi}\sigma_{{\rm m},i}}
e^{-x^{2}/2\sigma_{{\rm m},i}^{2}}
+\frac{P_{i}}{\sqrt{2\pi}\sigma_{{\rm t},i}} 
e^{-x^{2}/2\sigma_{{\rm t},i}^{2}}\,, \quad (i=1,2)\,.
\label{npdf1} 
\end{eqnarray}
The left panel of Fig.~\ref{fig:time_series} illustrates the probability 
distribution function of (\ref{npdf1}). 
This model can be characterized by the two parameters, i.e., 
the ratio of variance, $(\sigma_{{\rm t},i} / \sigma_{{\rm m},i})^{2}$ 
and the fraction of non-Gaussian tail, $P_{i}$. Here, $P_{i}$ means the 
total probability of the non-Gaussian tail. 
Of particularly interest is the case that 
$\sigma_{{\rm t},i} / \sigma_{{\rm m},i} > 1$ and $P_{i} \ll 1$. 
Thus, the detector noise is 
approximately described by the Gaussian distribution with the main 
variance $\sigma_{{\rm m},i}^{2}$, but to some extent, it exhibits the 
non-Gaussian tail  
characterized by the second component of the Gaussian distribution with 
a large variance $\sigma_{{\rm t},i}^{2}$. The examples of this situation 
are illustrated in the right panel of Fig.~\ref{fig:time_series}.

On the other hand, we assume that
the probability function of the stochastic signal is 
simply described by the Gaussian distribution with zero 
mean and with a small amplitude of the variance $\epsilon^2$: 
\begin{equation}
 p_{h^{k}}(h^{k})
=\frac{1}{\sqrt{2\pi}\epsilon} e^{-(h^{k})^{2}/2\epsilon^{2}}.
\label{spdf}
\end{equation}
Using these probability distribution functions, we derive the analytical formula 
for detection efficiency of the GCC statistic, 
i.e., $P_{\rm FA}$-$P_{\rm FD}$ curve and minimum detectable 
amplitude of the signal for gravitational-wave background.

\subsection{$P_{\rm FA}$ versus $P_{\rm FD}$ curve}
\label{subsec: PFA_PFD_curve}

In order to quantify the detection efficiency of the GCC statistic 
and compare it with that of the SCC statistic, it is convenient to 
compute the $P_{\rm FA}$-$P_{\rm FD}$ curve. 
For any detection statistic $\Lambda$, the false-alarm and the 
false-dismissal probabilities,  $P_{\rm FA}$ and $P_{\rm FD}$ 
are expressed as 
\begin{gather}
 P_{\rm FA}[\Lambda^{\ast}] = \int_{\Lambda^{\ast}}^{\infty} dx \,
 p_{\Lambda}^{(0)}(x), \notag \\
 P_{\rm FD}[\Lambda^{\ast}] =1- \int_{\Lambda^{\ast}}^{\infty} dx \, 
p_{\Lambda}^{(1)}(x). 
\label{pfapfd1}
\end{gather}
Here, $p_{\Lambda}^{(0)}(x)$ and $p_{\Lambda}^{(1)}(x)$ are 
the probability distributions of 
the decision statistic in the absence and the presence 
of the signal, respectively. Thus, the $P_{\rm FA}$-$P_{\rm FD}$ curve 
is simply obtained from Eq.(\ref{pfapfd1}) as the parametric function of 
the threshold $\Lambda^{\ast}$.  
According to the Neyman and Pearson criterion, the best strategy to
detect the stochastic signal is to choose the optimal statistic that 
minimizes the $P_{\rm FD}$ for a given value of $P_{\rm FA}$. 
In other words, if the $P_{\rm FD}$ of the GCC as function of $P_{\rm FA}$ 
is always smaller than that of the SCC, the GCC statistic 
is said to be more optimal compared to the SCC statistic.

In the large-sample limit ($N \gg 1$),
the central-limit theorem would be applicable and 
the probabilities $p_{\Lambda}^{(0)}$ and 
$p_{\Lambda}^{(1)}$ can be treated as a Gaussian function. 
We then have 
\begin{equation}
 p_{\Lambda}^{({\cal T})}(x)=
\frac{1}{\sqrt{2\pi}\Delta \Lambda^{({\cal T})}}
\exp\left[- \frac{(x- \langle \Lambda^{(\cal T)}\rangle)^{2}}
{2 (\Delta \Lambda^{({\cal T})})^{2}} \right] \,, \quad
({\cal T}=0,1)\,.
\label{pfapfd2}
\end{equation}
Here and in what follows, quantities
 $\langle \Lambda^{(0)}\rangle$ and $[\Delta \Lambda^{(0)}]^{2}$
denote the mean and the variance for a decision statistic 
in the absence of signal, while
$\langle \Lambda^{(1)}\rangle$ and $[\Delta \Lambda^{(1)}]^{2}$
are the mean and the variance for a decision statistic with a signal.
From Eqs.~(\ref{pfapfd1}) and (\ref{pfapfd2}), 
the $P_{\rm FA}$-$P_{\rm FD}$ curve is given by 
\begin{equation}
P_{\rm FD}=1-\frac{1}{2}{\rm erfc}\left[\left\{\sqrt{2}{\rm erfc}^{-1}
[2P_{\rm FA}]-\frac{\langle \Lambda^{(1)}\rangle}
{\Delta \Lambda^{(0)}}
+\frac{\langle \Lambda^{(0)}\rangle}{\Delta \Lambda^{(0)}}
\right\}
\frac{1}{\sqrt{2}}
\frac{\Delta \Lambda^{(0)}}
{\Delta \Lambda^{(1)}}
\right] .
\label{pfa-pfd_curve}
\end{equation}
Here, erfc$[x]$ is the complementary error function
defined by
\begin{equation}
{\rm {erfc}}[x]=\frac{2}{\sqrt{\pi}} \int_{x}^{\infty}dz \, e^{-z^{2}}.
\end{equation}
Note that in the case of the SCC statistic,
the quantity $\langle \Lambda^{(1)}\rangle / \Delta \Lambda^{(0)}$  
just coincides with the usual meaning of the signal-to-noise ratio (SNR).
In general, the false-dismissal probability 
$P_{\rm FD}$ is a decreasing function of the quantity
$\langle \Lambda^{(1)}\rangle /
\Delta \Lambda^{(0)}$ for a given probability of false alarm $P_{\rm FA}$.

\subsection{Mean and Variance for detection statistic}
\label{subsec: mean_var_GCC}
Next,

Our task is to calculate the means and the variances for the detection 
statistic, i.e., $\langle \Lambda^{({\cal T})} \rangle $ and 
$\left[ \Delta\Lambda^{({\cal T})}\right]^2$.  
In order to compare the performance of the GCC statistic to that of the
SCC statistic, we first consider the means and the variances for
the SCC statistic. From Eqs.~(\ref{output}) 
and (\ref{scc})-(\ref{spdf}), the ensemble averages become
\begin{eqnarray}
\langle \Lambda_{\rm SCC}^{(0)} \rangle &=& 0, 
\\
\left[ \Delta\Lambda_{\rm SCC}^{(0)} \right]^{2} 
& = &
\Big\langle
\left[ \Lambda_{\rm SCC}^{(0)} - {\big\langle \Lambda_{\rm SCC}^{(0)} \big\rangle}
\right]^{2} 
\Big\rangle
 = \frac{\langle n_{1}^{2}\rangle \langle n_{2}^{2} \rangle}{{N}}, 
\label{scc0variance} \\
\langle \Lambda_{\rm SCC}^{(1)} \rangle &=& \epsilon^{2}, 
\\
\left[ \Delta\Lambda_{\rm SCC}^{(1)} \right]^{2} 
& = &
\Big\langle
\left[ \Lambda_{\rm SCC}^{(1)} - 
    \big\langle  \Lambda_{\rm SCC}^{(1)} \big\rangle
 \right]^{2} 
\Big\rangle
 =  \frac{1}{{N}}\left[
2\langle\Lambda_{\rm SCC}^{(1)}\rangle^{2}+
\langle\Lambda_{\rm SCC}^{(1)}\rangle(\langle n_{1}^{2}\rangle
+\langle n_{2}^{2} \rangle)+
\langle n_{1}^{2} \rangle \langle n_{2}^{2} \rangle \right],
\label{scc1variance}
\end{eqnarray}
where
\begin{equation}
\left[ \Delta\Lambda_{\rm SCC}^{({\cal T})} \right]^{2}
\equiv 
\Big\langle
    \left[ \Lambda_{\rm SCC}^{({\cal T})}-
\big\langle  \Lambda_{\rm SCC}^{({\cal T})} \big\rangle 
\right]^{2} 
\Big\rangle
\quad {\rm and} \quad
\langle n_{i}^{2} \rangle = (1-P_{i})\sigma_{{\rm m},i}^{2}+
P_{i}\sigma_{{\rm t},i}^{2} \, . \,
\label{definition1} 
\end{equation}

Next, we calculate the means and the variances 
for the GCC statistic (\ref{gcc}). 
For the non-Gaussian model (\ref{npdf1}) of the instrumental noises, 
the derivative $f_{i}'(x)$ in  Eq.~(\ref{gcc}) is given by 
\begin{equation}
f_i'(x)=\frac{x}{\sigma_{{\rm m},i}^{2}}\left[ \frac{(1-P_i)+
P_i(\sigma_{{\rm m},i} / \sigma_{{\rm t},i})^{3}
e^{x^{2}(\sigma_{{\rm m},i}^{-2}-{\sigma_{{\rm t},i}^{-2}})/2}}
{(1-P_i)+P_i(\sigma_{{\rm m},i} /\sigma_{{\rm t},i} )
e^{x^{2}(\sigma_{{\rm m},i}^{-2}-\sigma_{{\rm t},i}^{-2})/2}}  
\right].
\label{eq:derivative_f}
\end{equation}
The expression (\ref{eq:derivative_f}) seems rather intractable to 
further develop the analytical calculation.  However, 
in the situations we are interested in, i.e.,  
$\sigma_{{\rm t},i} / \sigma_{{\rm m},i}>1$ and $P_{i}\ll1$, 
the above function simply behaves like 
$f_i'(x) \approx x/\sigma_{{\rm m},i}^{2}$ for small value of $|x|$ 
and $f_i'(x) \approx x/\sigma_{{\rm t},i}^{2}$ for large value of $|x|$. 
Thus,
one may apply the two-step approximation to the 
function (\ref{eq:derivative_f}) as:
\begin{equation}
f_{i}'(s_{i}^{k})
\equiv  \begin{cases} 
\quad  s_{i}^{k} 
& \text{$ ; \lvert s_{i}^{k} \rvert \leq  \lvert x_{{\rm cr},i} \rvert  $\,,} 
\\
\left( \frac{\sigma_{{\rm m},i}}{\sigma_{{\rm t},i}}\right)^{2} \, s_{i}^{k}
& \text{$ ; \lvert s_{i}^{k} \rvert > \lvert x_{{\rm cr},i} \rvert $}\,.
\end{cases}
\label{twostep}
\end{equation}
Here, the quantity $x_{{\rm cr},i}$ is the critical value that  
characterizes the boundary between small $\lvert s_{i}^{k} \rvert$ and 
large $\lvert s_{i}^{k} \rvert$. Note that we adjust
the overall factor of the function $f_{i}'(s_{i}^{k})$
so as to coincide with the SCC
statistic (\ref{scc})
in the limit $\lvert x_{{\rm cr},i} \rvert \rightarrow \infty$.

\begin{figure}[!tbp]
\begin{center}
\includegraphics[width=8cm,angle=0,clip]{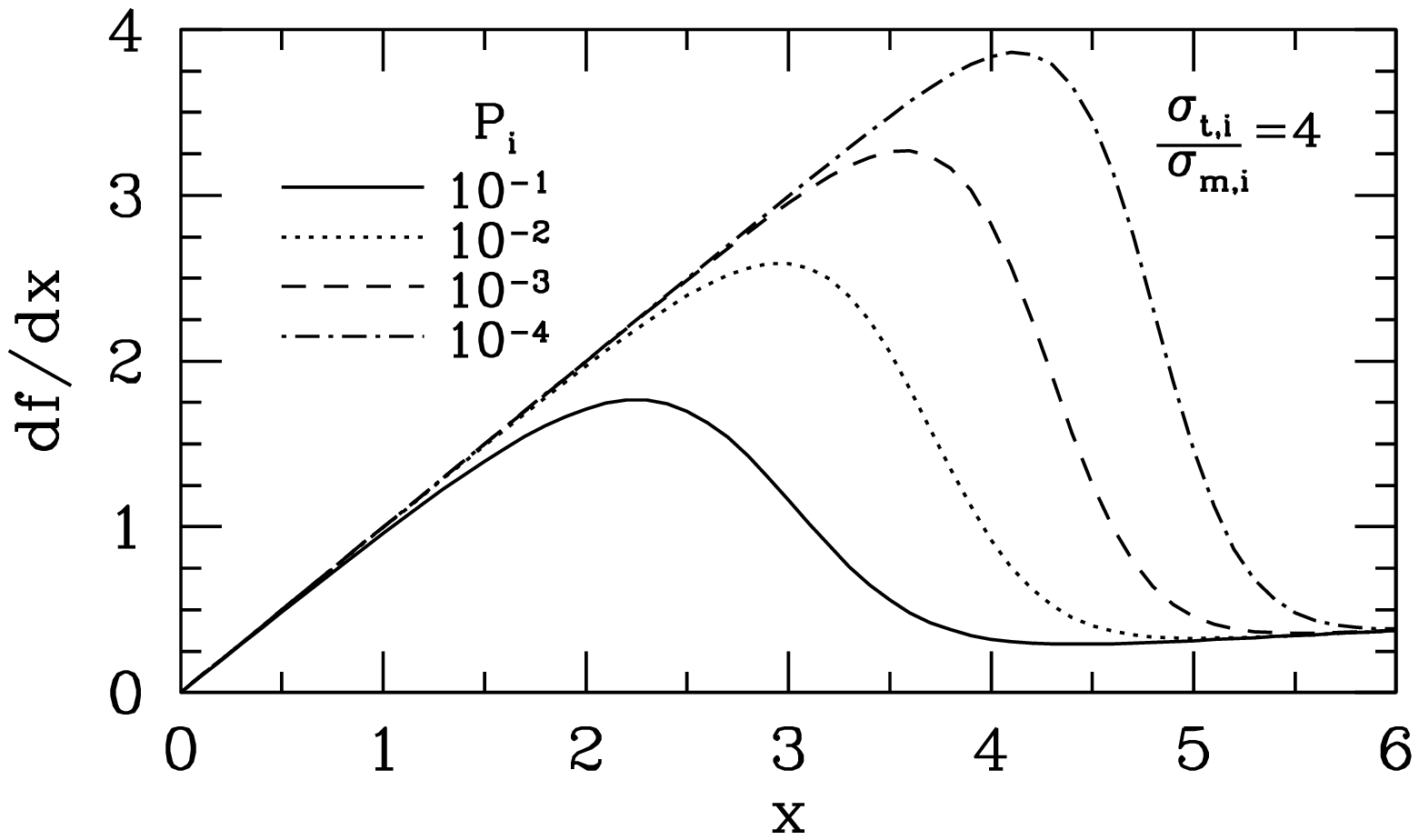}
\hspace*{0.5cm}
\includegraphics[width=8cm,angle=0,clip]{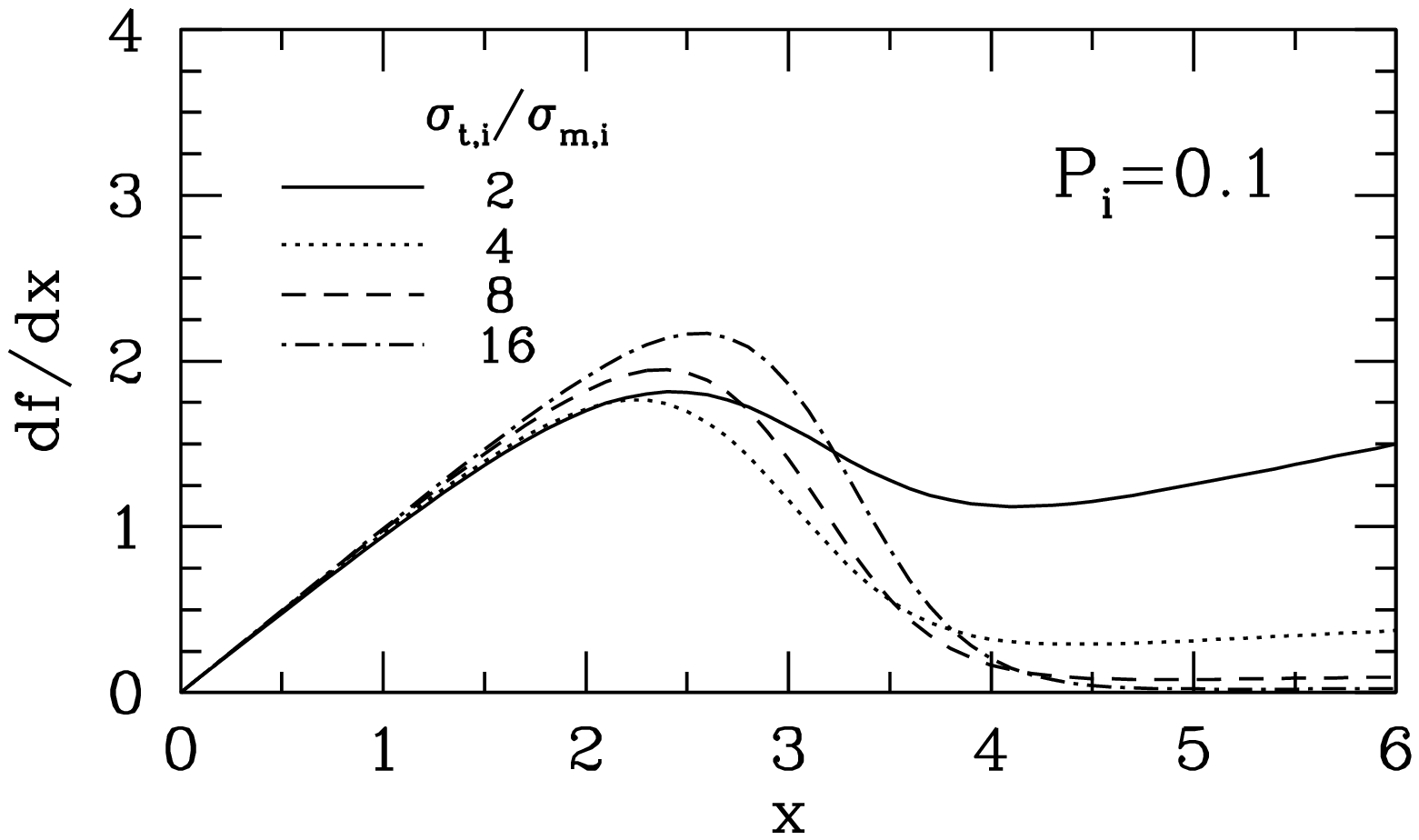}
\end{center}
    \caption{Derivative of the function $f(x)$ in the two-component
 Gaussian noise model. Left panel shows  
 the dependence of the tail fraction $P_i$ keeping the 
 ratio of noise variance fixed, i.e., 
$\sigma_{{\rm t},i}/\sigma_{{\rm m},i}=4$. On the other hand, right 
panel presents the dependence of the ratio 
$\sigma_{{\rm t},i}/\sigma_{{\rm m},i}$ keeping 
the tail fraction fixed, i.e., $P_i=0.1$. }
 \label{fig:critical_value}
\end{figure}

Fig.~\ref{fig:critical_value} 
shows the dependence of the function $f'(x)$ on the model parameters 
$P_i$ ({\it left}) and $\sigma_{{\rm t},i} / \sigma_{{\rm m},i}$ 
({\it right}). As decreasing the tail fraction or increasing
the ratio $\sigma_{{\rm t},i} / \sigma_{{\rm m},i}$, 
the asymptotic behavior of $f'(x)$ steeply changes 
from $x/\sigma_{{\rm m},i}^{2}$ to 
$x/\sigma_{{\rm t},i}^{2}$ around the inflection point of $f'(x)$. 
Hence, it seems reasonable to set the critical value $x_{{\rm cr},i}$ to 
the inflection point of $f'(x)$. Then,   
the quantity $x_{{\rm cr},i}$ is approximately expressed as 
\begin{equation}
x_{{\rm cr},i}^{2} \approx
\frac{\sigma_{{\rm m},i}^{2} 
\sigma_{{\rm t},i}^{2}}{\sigma_{{\rm t},i}^{2}-\sigma_{{\rm m},i}^{2}}
\left(\sqrt{12+\left(\log\left[\frac{P_{i}}{1-P_{i}} 
\frac{\sigma_{{\rm m},i}}{\sigma_{{\rm t},i}}\right]\right)^{2}}
-\log\left[\frac{P_{i}}{1-P_{i}} 
\frac{\sigma_{{\rm m},i}}{\sigma_{{\rm t},i}}\right]\right).
\label{critical}
\end{equation}
Here, we only considered the solution
satisfying the condition $(x_{{\rm cr},i}/\sigma_{{\rm m},i})>1$.

Adopting the critical value Eq.~(\ref{critical}), with a help of
two-step approximation, 
the means and the variances of the GCC statistic can be analytically 
calculated.  The details of the calculation are presented in 
Appendix~\ref{appen:appendix1}.  The resultant expressions become 
\begin{eqnarray}
\langle \Lambda_{\rm GCC}^{(0)} \rangle &=& 0\,, \label{gcc0meantrue}\\
\Delta\Lambda_{\rm GCC}^{(0)}
&=& \frac{1}{\sqrt{N}}[\langle n_{1}^{2} \rangle_{\rm G} \langle n_{2}^{2}\rangle_{\rm G} ]^{1/2}\,, 
\label{gcc0variancetrue} \\
\langle \Lambda_{\rm GCC}^{(1)} \rangle & = & 
\{1-(P_{1}+P_{2})\} \,  
P_{\rm G}[x_{{\rm cr},1},\sigma_{{\rm m},1}]\, 
P_{\rm G}[x_{{\rm cr},2},\sigma_{{\rm m},2}]\, \epsilon^{2} \,
+\, {\rm higher \, order \, terms}\,,
\label{gcc1mean} \\
\Delta\Lambda_{\rm GCC}^{(1)} 
 & =  & 
\frac{1}{\sqrt{N}} \left[\langle n_{1}^{2} \rangle_{\rm G}\langle n_{2}^{2} 
 \rangle_{\rm G}+ \mathcal{O}(\epsilon^{2} \cdot \langle n_{i}^{2} \rangle_{\rm G})
\right]^{1/2}\,,
\label{gcc1variancetrue}
\end{eqnarray}
where we defined 
\begin{eqnarray}
&&P_{{\rm G}}[x,\sigma] \equiv  
   {\rm erf}
\left[\frac{x}{\sqrt{2}\sigma} \right]  
-\sqrt{\frac{2}{\pi}}
\frac{x}{\sigma} e^{
-(x/\sigma)^{2}/2} \,,  \label{define1}\\
\langle n_{i}^{2} \rangle_{\rm G} 
&\equiv&(1-P_{i}) \, \sigma_{{\rm m},i}^{2} \,
P_{{\rm G}}[x_{{\rm cr},i},\sigma_{{\rm m},i}]
+ P_{i} \, \sigma_{{\rm t},i}^{2} \,
P_{{\rm G}}[x_{{\rm cr},i},\sigma_{{\rm t},i}] \,. \label{noise_mean}  
\end{eqnarray}
Here the quantity erf$[x]$ is the error function. In deriving
Eqs.(\ref{gcc0meantrue}-\ref{gcc1variancetrue}), we have neglected 
contributions 
of the integral from the region $[x_{{\rm cr},i},\infty]$.  
In Ref.~\cite{Allen:2001ay}, this treatment is called {\it clipping}.
The explicit expressions of 
the higher-order terms in Eq.(\ref{gcc1mean})
are given in Appendix~\ref{appen:appendix1}. These terms 
turn out to be subdominant
if the non-Gaussian parameters
become $P_{i} \lesssim 0.2$ or $\sigma_{{\rm t},i}/\sigma_{{\rm m},i}
\gtrsim 3$. In what follows,
we neglect the higher-order 
terms in Eq.(\ref{gcc1mean}) unless otherwise stated.

Now, we substitute the expressions 
Eqs.~(\ref{critical})-(\ref{gcc1variancetrue}) 
into Eq.(\ref{pfa-pfd_curve}). The analytic 
$P_{\rm FA}$-$P_{\rm FD}$ curve for the GCC statistic is 
written as 
\begin{equation}
P_{\rm FD}=1-\frac{1}{2}{\rm erfc}\left[\left\{\sqrt{2}{\rm erfc}^{-1}
[2P_{\rm FA}]-\rho\left(\frac{\rho_{\rm eff}}{\rho}\right)
\right\}\frac{\Delta}{\sqrt{2}}
\left(\frac{\Delta_{\rm eff}}{\Delta}\right)\right],
\label{eq:ana_pfa-pfd_curve}
\end{equation}
where
\begin{equation}
 \rho =\frac{\langle \Lambda_{\rm SCC}^{(1)}\rangle}
{\Delta \Lambda_{\rm SCC}^{(0)}}, \quad 
\Delta =\frac{\Delta \Lambda_{\rm SCC}^{(0)}}
{\Delta \Lambda_{\rm SCC}^{(1)}}, \quad
 \rho_{\rm eff} =\frac{\langle \Lambda_{\rm GCC}^{(1)}\rangle}
{\Delta \Lambda_{\rm GCC}^{(0)}}, \quad 
\Delta_{\rm eff} =\frac{\Delta \Lambda_{\rm GCC}^{(0)}}
{\Delta \Lambda_{\rm GCC}^{(1)}}.
\end{equation}
In the expression (\ref{eq:ana_pfa-pfd_curve}), 
we have introduced the auxiliary quantities $\rho$ and $\Delta$ 
to clarify the differences between the GCC and the SCC statistics. 
Obviously, the ratios $\rho_{\rm eff}/\rho$ and 
$\Delta_{\rm eff}/\Delta$ become unity when the probability 
distribution of noises is Gaussian, leading to the $P_{\rm FA}$-$P_{\rm FD}$ 
curve for SCC statistic. Thus, the deviation of these 
quantities from unity characterizes the efficiency of the GCC statistic.

\begin{figure}[!tbp]
\begin{center}
\includegraphics[width=8cm,angle=0,clip]{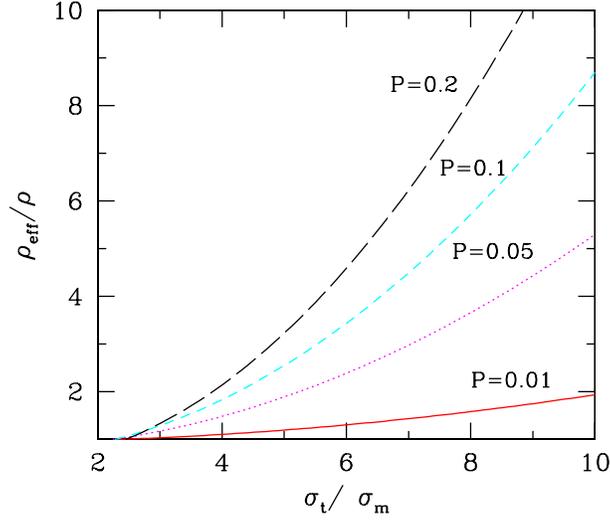}
\end{center}
    \caption{The quantity
$\rho_{\rm eff}/\rho$ defined in Eq.(\ref{eq:ana_pfa-pfd_curve}) 
as function of $\sigma_{\rm t}/\sigma_{\rm m}$  
in the case of the two identical detectors. 
From top to bottom, the tail fraction $P$ is chosen as 
$P=0.2$, $0.1$, $0.05$ and $0.01$. 
}
\label{fig:rho_ratio}
\end{figure}

Fig.~\ref{fig:rho_ratio} shows the ratio $\rho_{\rm eff}/\rho$ as 
the function of $\sigma_{{\rm t}}/\sigma_{{\rm m}}$ 
for various tail fraction $P$. 
To plot the curves, just for simplicity, we assume that two detectors 
are identical:
\begin{equation}
\sigma_{\rm m}\equiv \sigma_{{\rm m},1}=\sigma_{{\rm m},2}, \quad
\sigma_{\rm t}\equiv \sigma_{{\rm t},1}=\sigma_{{\rm t},2} , \quad
P\equiv P_{1}=P_{2}, \quad 
x_{{\rm cr}}\equiv x_{{\rm cr},1}=x_{{\rm cr},2}. 
\label{identical}
\end{equation}
In Fig.~\ref{fig:rho_ratio},  the ratio $\rho_{\rm eff}/\rho$ is 
always larger than unity for any values of $P$ and 
$\sigma_{\rm t}/\sigma_{\rm m}$. 
Recall that the quantity $\rho$ has the usual meaning of the SNR, 
this result implies that the clipping taken in the GCC statistic 
always leads to a larger effective SNR than that of the SCC statistic. 
On the other hand, when we evaluate the quantity 
$\Delta_{\rm eff}/\Delta$,  
one finds that this ratio is always less than $1$. 
These two facts indicate that 
the false-dismissal probability $P_{\rm FD}$ of the GCC statistic 
is always smaller than that of the SCC statistic. 
Note also that  
$\Delta_{\rm eff} \approx 1 $ and $\Delta\approx 1$
as long as 
the signal $\epsilon$ is small. 
Thus, for a good approximation, 
we can set 
$\Delta_{\rm eff}$ to unity.
Hence, the performance of the GCC statistic is mainly 
attributed to the ratio $\rho_{\rm eff}/\rho$.

\begin{figure}[!tbp]
\begin{center}
\includegraphics[width=8cm,angle=0,clip]{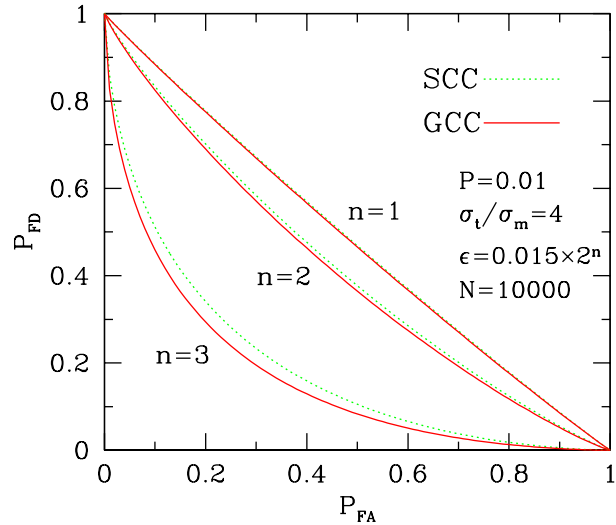}
\end{center}
    \caption{Analytic $P_{\rm FA}$-$P_{\rm FD}$ curves 
for the standard and generalized
cross-correlation statistics in the presence of the non-Gaussian noises 
described by the specific model (\ref{npdf1}).  
The sold (dashed) lines represent the 
$P_{\rm FA}$-$P_{\rm FD}$ curves for the GCC (SCC) statistic for 
the stochastic signals with amplitude $\epsilon=0.03$, $0.06$ and $0.12$ 
(top to bottom). Here, we assume that the two detectors are identical 
(see Eq.~(\ref{identical})). For each curves, the parameters are set as
$P=0.01$, $\sigma_{\rm t}/\sigma_{\rm m}=4$ and $N=10^4$. 
}   
     \label{fig:ana_pfa-pfd_curve}
\end{figure}

Based on this consideration, in Fig.~\ref{fig:ana_pfa-pfd_curve}, 
we plot the analytic $P_{\rm FA}$-$P_{\rm FD}$ curves 
for various signal amplitudes. Here, the parameters $P$,  
$\sigma_{\rm t}/\sigma_{\rm m}$ and $N$ are specifically chosen to 
$P=0.01$, $\sigma_{\rm t}/\sigma_{\rm m}=4$ and $N=10^4$. 
The solid and dotted lines 
represent the $P_{\rm FA}$-$P_{\rm FD}$ curves for the GCC and the SCC 
statistics, respectively. In each signal amplitude $\epsilon$, 
the false-dismissal probability $P_{\rm FD}$ of the GCC statistic 
is always smaller than that of the SCC statistic for any $P_{\rm FA}$. 
As expected, the performance of the GCC statistic improves 
as the parameter $\epsilon$ increases.


\subsection{Minimum detectable amplitude}
\label{subsec:epsilon_min}

In addition to the $P_{\rm FA}$-$P_{\rm FD}$ 
curves, the minimum detectable amplitude of the stochastic signal,  
$\epsilon_{\rm detect}$ is a direct 
measure to quantify the performance of the detectability. 
In order to estimate this statistically, we must
first specify the threshold values $(P_{\rm FA}^{*},~P_{\rm FD}^{*})$ 
called detection point \cite{Drasco:2002yd}. 
For given threshold values, the minimum detectable
amplitude $\epsilon_{\rm detect}$ can be uniquely determined 
from Eq.(\ref{eq:ana_pfa-pfd_curve}). 
For simplicity, we set
$P_{\rm FA}^{*}$=$P_{\rm FD}^{*}$. The resultant amplitude for the GCC 
statistic,  $\epsilon_{\rm detect}^{\rm GCC}$ is 
\begin{eqnarray}
\{\epsilon_{\rm detect}^{\rm GCC}\}^2&=&\frac{\displaystyle \left\{
\langle n_{1}^{2} \rangle_{\rm G} \langle n_{2}^{2}\rangle_{\rm G} 
\right\}^{1/2}}{\displaystyle \sqrt{N}}\,\,
\frac{\displaystyle 2\sqrt{2}\gamma
}{\displaystyle \{1-(P_{1}+P_{2})\} \,  
P_{\rm G}[x_{{\rm cr},1},\sigma_{{\rm m},1}]\, 
P_{\rm G}[x_{{\rm cr},2},\sigma_{{\rm m},2}]} ,
\notag \\ 
&=& 
G^2(\sigma_{{\rm m},1},\sigma_{{\rm m},2},\sigma_{{\rm t},1},
\sigma_{{\rm t},2},P_{1},P_{2}) \,\,\{\epsilon_{\rm detect}^{\rm SCC}\}^2 \, ,
\label{minimum_epsilon}
\end{eqnarray}
where we have assumed $\Delta_{\rm eff}=1$.
The quantity $\gamma$ is given by 
$\gamma = {\rm erfc}^{-1}[2P_{\rm FA}^{*}]$ and 
the amplitude $\epsilon_{\rm detect}^{\rm SCC}$ means the 
minimum detectable amplitude for the SCC statistic 
in the large $N$ limit \cite{Drasco:2002yd}: 
\begin{eqnarray}
\{\epsilon_{\rm detect}^{\rm SCC}\}^2=
\frac{2\sqrt{2} \gamma 
\left\{\langle n_{1}^{2} \rangle \langle n_{2}^{2}\rangle 
\right\}^{1/2}}{\sqrt{N}}.  
\label{minimum_epsilon_SCC}
\end{eqnarray}
In Eq. (\ref{minimum_epsilon}), the important quantity is 
the function $G$ characterizing the gain compared to the amplitude 
$\epsilon_{\rm detect}^{\rm SCC}$:  
\begin{eqnarray}
G^2(\sigma_{{\rm m},1},\sigma_{{\rm m},2},
\sigma_{{\rm t},1},\sigma_{{\rm t},2},P_{1},P_{2}) 
\equiv \frac{1}{\{1-(P_{1}+P_{2})\} \,  
P_{\rm G}[x_{{\rm cr},1},\sigma_{{\rm m},1}]\, 
P_{\rm G}[x_{{\rm cr},2},\sigma_{{\rm m},2}]}
\left(\frac
{\langle n_{1}^{2} \rangle_{\rm G}
\langle n_{2}^{2} \rangle_{\rm G}}
{\langle n_{1}^{2} \rangle \langle n_{2}^{2}\rangle }
\right)^{\frac{1}{2}} \, .
\label{gain}
\end{eqnarray}
The function $G$ becomes unity when the noise probability functions 
reduce to the Gaussian distribution. It also 
approaches unity if the ratio of the noise variance 
$\sigma_{\rm t,i}/\sigma_{\rm m,i}$ becomes unity. 
For the stochastic signal $\epsilon$, we have the relation
$\Omega_{\rm gw} \propto \epsilon^{2} \propto$ SNR.
Thus, the minimum detectable $\Omega_{\rm gw}$ using the GCC statistic
is improved by a factor $G^{2}$, compared to that of the SCC statistic.

\begin{figure}[!]
\begin{center}
\includegraphics[width=8cm,angle=0,clip]{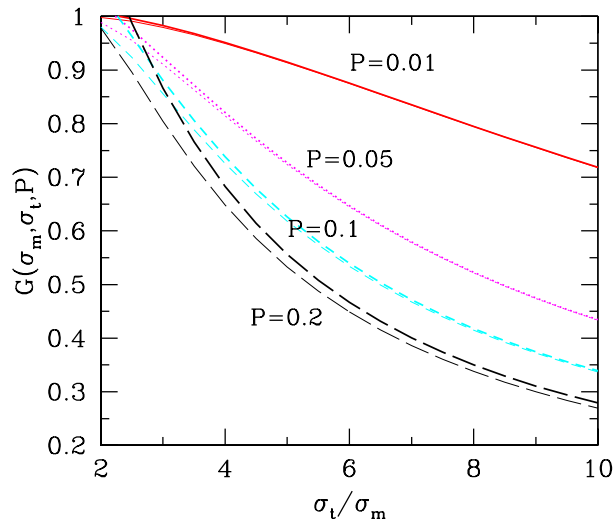}
\end{center}
\caption{
The function $G$ plotted against the ratio  
$\sigma_{\rm t}/\sigma_{\rm m}$ in the case of two
 identical detectors (see Eq.~(\ref{identical})). Here, the 
tail fraction $P$ is specifically chosen as 
$P=0.01$, $0.05$, $0.1$ and $0.2$ from top to bottom.
The thick and thin lines are the function $G$ defined in Eq.~(\ref{gain}) and 
the one taking account of the higher-order terms (\ref{higher}), 
respectively. }

    \label{fig:gain_factor}
\end{figure}

In Fig.~\ref{fig:gain_factor},
 the thick lines show the quantity $G$  
as function of $\sigma_{\rm t}/\sigma_{\rm m}$ 
in the case of two identical detectors 
(see Eq.~(\ref{identical})). The thin lines
represent the same plot, but we have taken account of
the higher-order terms
(\ref{higher}) in Appendix~\ref{appen:appendix1}.
As the tail fraction becomes smaller and the ratio 
$\sigma_{\rm t}/\sigma_{\rm m}$ becomes larger, the thick lines
tend to approach thin lines. 
The quantity $G$ monotonically decreases as increasing 
the ratio $\sigma_{\rm t}/\sigma_{\rm m}$ or the tail fraction $P$.  
Specifically, for the parameters 
$P=0.1$ and $\sigma_{\rm t}/\sigma_{\rm m}=10$, 
we obtain $G\sim 0.35$. This implies that the sensitivity to the stochastic 
signal is improved by a factor $10$ in terms of SNR, 
compared to the sensitivity achieved 
with the SCC statistic. 

In the situation with $P_{i} \ll 1$ and 
$(\sigma_{{\rm m},i}/\sigma_{{\rm t},i}) \lesssim 1$,
a more compact form of the approximation for $G^{2}$ 
is found :
\begin{equation}
G^2 \simeq 
\left(\frac
{\langle n_{1}^{2} \rangle_{\rm G}
\langle n_{2}^{2} \rangle_{\rm G}}
{\langle n_{1}^{2} \rangle \langle n_{2}^{2}\rangle }
\right)^{\frac{1}{2}} \simeq
\prod_{i=1,2} 
\left[\frac{1-P_i}{(1-P_i)+P_i\,\, 
(\sigma_{\rm t,i}/\sigma_{\rm m,i})^2}\right]^{1/2}\,. 
\label{eq:approx_G}
\end{equation}
Thus, 
when the quantity $P_i\,(\sigma_{\rm t,i}/\sigma_{\rm m,i})^2$ is 
larger than unity,  the GCC statistic can become more powerful 
than the SCC statistic.

\section{Monte Carlo simulation}
\label{sec:4}

In this section, we perform Monte Carlo simulations of 
the cross-correlation analysis and compare 
the $P_{\rm FA}$-$P_{\rm FD}$ curves and 
the minimum amplitude $\epsilon_{\rm detect}$ from  
the analytic estimates 
with those obtained from the numerical simulations. 
For the rest of this paper, we specifically assume that the 
two detectors are identical and satisfy the condition (\ref{identical}).

\subsection{Algorithm of Monte Carlo simulation}

Our Monte Carlo algorithm basically follows 
Ref.\cite{Drasco:2002yd}. 
We numerically calculate the false-alarm and false-dismissal
probabilities $P_{\rm FA}$ and $P_{\rm FD}$ 
by conducting an ensemble over the $N_{\rm\scriptscriptstyle CHUNK}$ 
simulated experiments. 
For each experiment, we randomly generate two kinds of
$(N \times 2)$ matrix ${\cal S}$ 
made up of the detector outputs, in which one output contains 
stochastic signals and other data contain only the instrumental noises.  
We then compute the
decision statistic in the presence or the absence of the stochastic signals.  
Choosing the threshold for the decision statistic, we obtain 
$P_{\rm FA}$-$P_{\rm FD}$ curve. The details of the algorithm 
are summarized as follows (see also Ref.\cite{Drasco:2002yd}):
\begin{itemize}
\item{
Generate two kind of  $(N \times 2)$ data matrix ${\cal S}$ :} \\
For a specific parameter set $(P,\sigma_{\rm m},\sigma_{\rm t},\epsilon,N)$, 
we first generate the $N$ data train 
which only contains the instrumental noises, i.e., 
$s_i^k=n_k^i$ $(i=1,2,~k=1,\cdots,N)$. These random data are 
created according to the probability distribution function (\ref{npdf1}). 
We then duplicate the data train and further add the stochastic signals 
(Eq.~(\ref{spdf})),  
to the one data train, i.e., $s_i^k=h_k^i+n_k^i$ $(i=1,2,~k=1,\cdots,N)$. 
\item{Compute the decision statistics $\Lambda_{\rm GCC}^{({\cal T})}$ 
      and $\Lambda_{\rm SCC}^{({\cal T})}$ 
      from the matrix ${\cal S}$ for ${\cal T}=0$ and $1$:} \\
Based on the expressions (\ref{gcc})
and (\ref{scc}), under a prior knowledge 
of the noise parameters $(P,\sigma_{\rm m},\sigma_{\rm t})$,   
we compute the decision statistics $\Lambda_{\rm GCC}^{({\cal T})}$ and 
$\Lambda_{\rm SCC}^{({\cal T})}$ from the data matrix ${\cal S}$ 
in both absence and presence of the 
stochastic signals $({\cal T}=0, 1)$.
Note that the derivative $f_{i}'(x)$ in  Eq.~(\ref{gcc}) is given by
     Eq.(\ref{eq:derivative_f}).

\item{Set a threshold value $\Lambda^{\ast}$ 
to determine a point $(P_{\rm FA}[\Lambda^{\ast}] ,
P_{\rm FD}[\Lambda^{\ast}])$ for GCC and SCC:}\\   
For a given value $\Lambda^{\ast}$, we increase $P_{\rm FA}$ 
by the factor $1/N_{\rm\scriptscriptstyle CHUNK}$ when 
the condition $\Lambda^{(0)} > \Lambda^{\ast}$ is satisfied. 
Also, we increase $P_{\rm FD}$ 
by $1/N_{\rm\scriptscriptstyle CHUNK}$ if the relation 
$\Lambda^{(1)} < \Lambda^{*}$ holds. 
These operations are performed in each case of the GCC and the SCC 
statistics by varying the threshold value $\Lambda^{\ast}$. 
\item{Repeat the above steps $N_{\rm\scriptscriptstyle CHUNK}$ times to 
    estimate the probabilities $(P_{\rm FA}[\Lambda^{\ast}] ,
P_{\rm FD}[\Lambda^{\ast}])$ } for various threshold values $\Lambda^{\ast}$. 
 \end{itemize}

In the simulations presented below, 
the numbers of samples and trials are set to $N=10^{4}$ and 
$N_{\rm\scriptscriptstyle CHUNK}=5 \times 10^{3}$, respectively. 
Note that the $N=10^4$ samples roughly correspond to the 
data points appropriate for the low-frequency detector like 
Laser Interferometer Space Antenna (LISA) \cite{Bender:1998}, 
for which $1$ year observation 
and the effective bandwidth $10^{-3}$Hz are assumed.  
Below, we will present the results under keeping 
the noise variance $\sigma_{\rm m}^{2}=1$ fixed.

\subsection{Simulation results and discussion}

\begin{figure}[ht]
\begin{center}
\includegraphics[width=8cm,angle=0,clip]{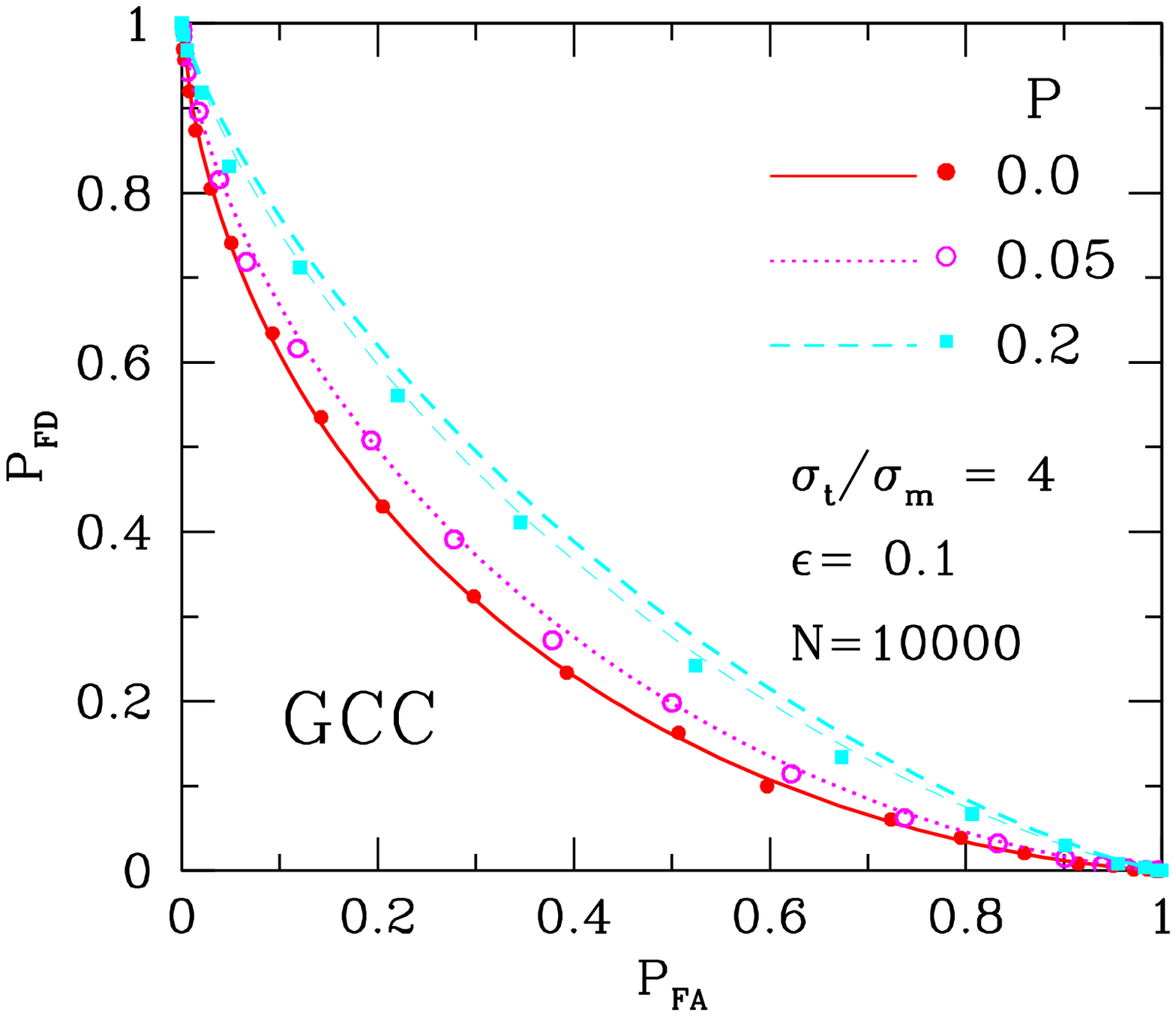}
\hspace*{0.5cm}
\includegraphics[width=8cm,angle=0,clip]{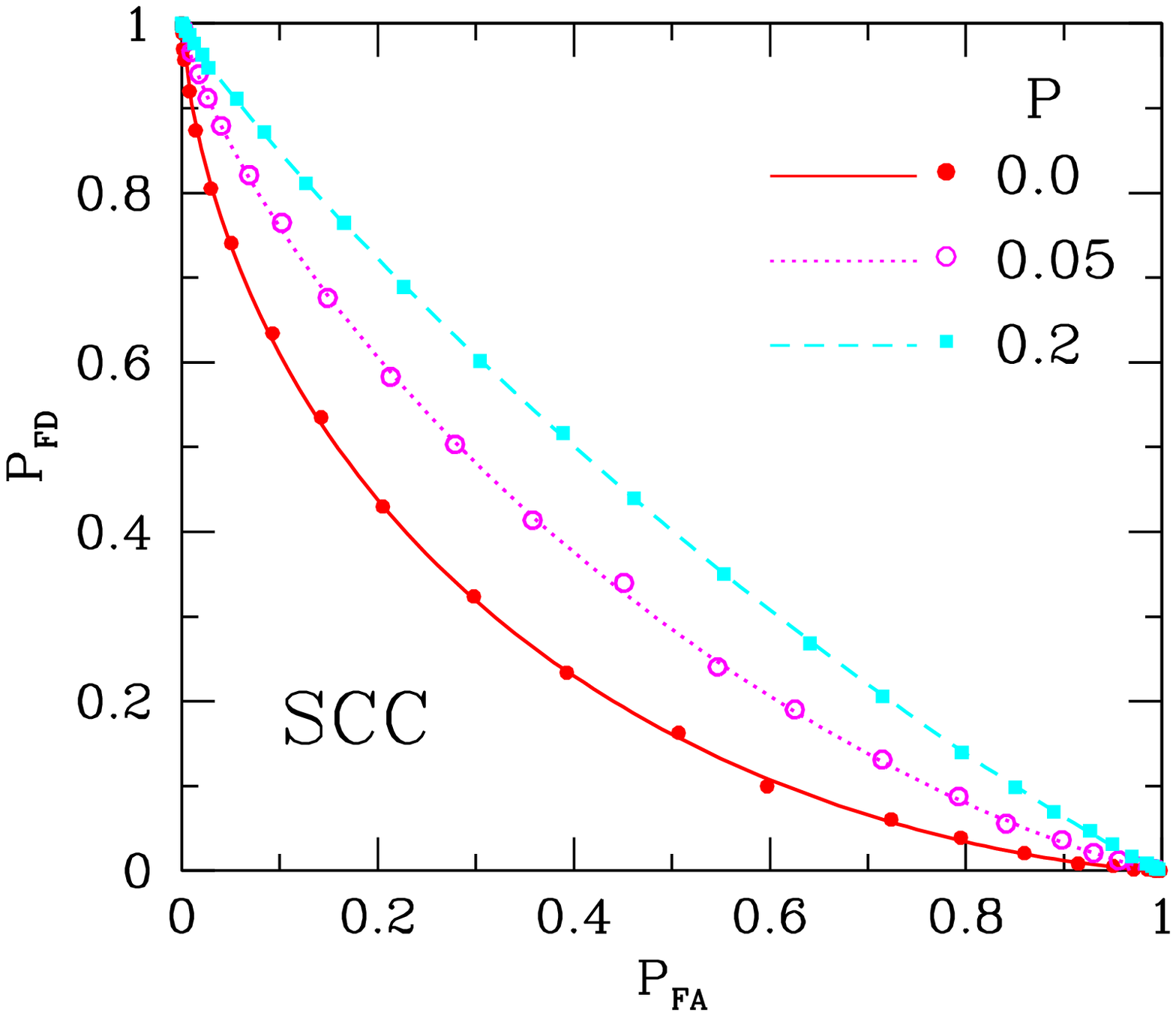}
\end{center}
    \caption{
$P_{\rm FA}$-$P_{\rm FD}$ curves  for the GCC ({\it left}) 
and the SCC ({\it right}) statistics.  Symbols denote the 
simulation results, while the lines indicate the 
analytic prediction from Eq.~(\ref{eq:ana_pfa-pfd_curve}). 
In each panel, the ratio of the noise variance is fixed to 
$\sigma_{\rm t}/\sigma_{\rm m}=4$ and the amplitude of stochastic 
signal is set to $\epsilon=0.1$. 
Note that for the tail fraction $P=0.0$, corresponding to the 
Gaussian noise case, the solid line and the filled circles
in left panel are identical to the one in right panel: 
$P=0.0$({\it filled circles} and {\it solid}); $P=0.05$ 
({\it open circles} and {\it dotted}); $P=0.2$ ({\it filled squares} and 
{\it dashed}). The thin dashed line for $P=0.2$ indicates
the analytic $P_{\rm FA}$-$P_{\rm FD}$ curve taking account of
the higher-order terms
(\ref{higher}).}
     \label{fig:pfa-pfd_curve_vary_P}
\end{figure}
\begin{figure}[ht]
\begin{center}
\includegraphics[width=8cm,angle=0,clip]{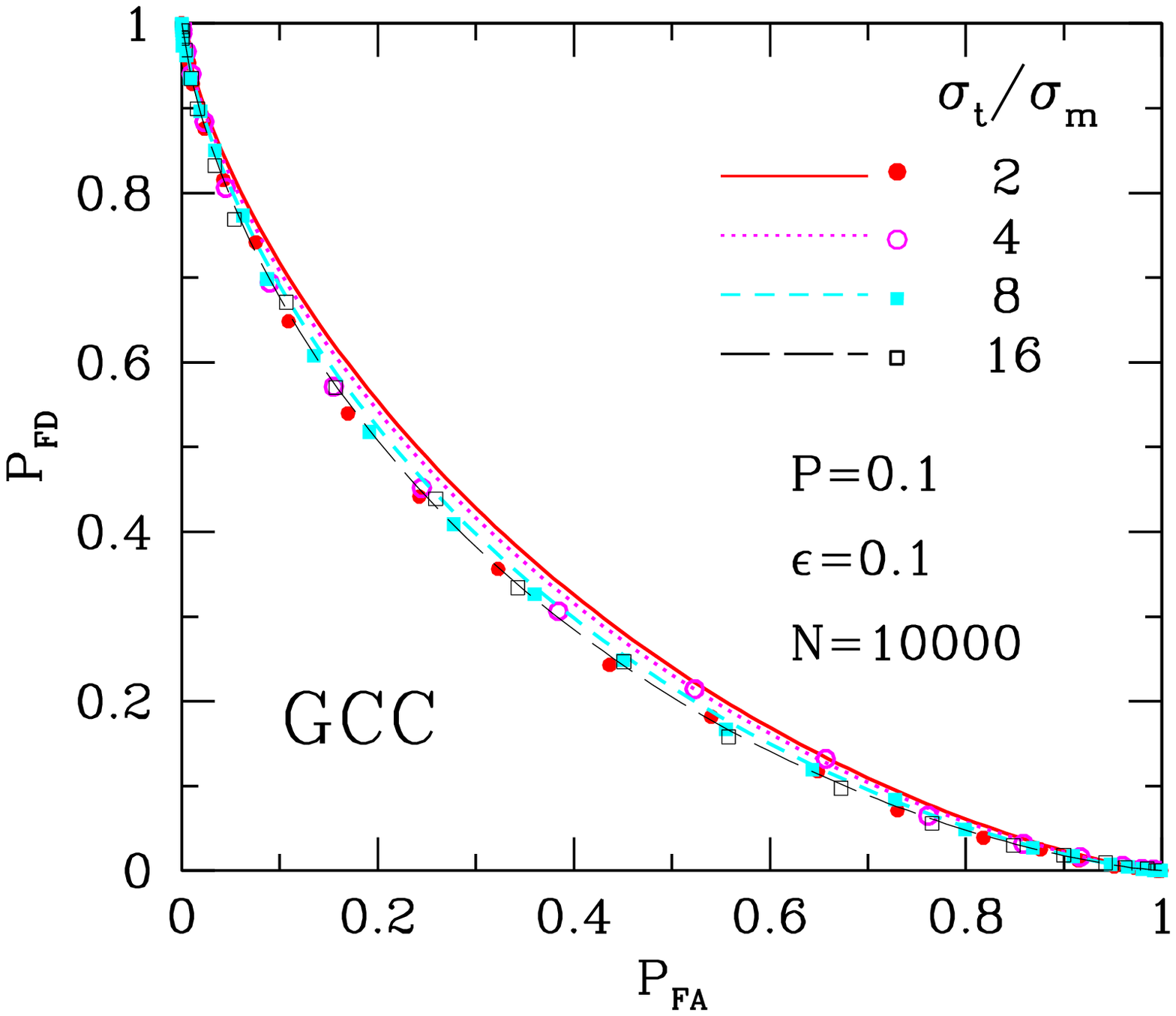}
\hspace*{0.5cm}
\includegraphics[width=8cm,angle=0,clip]{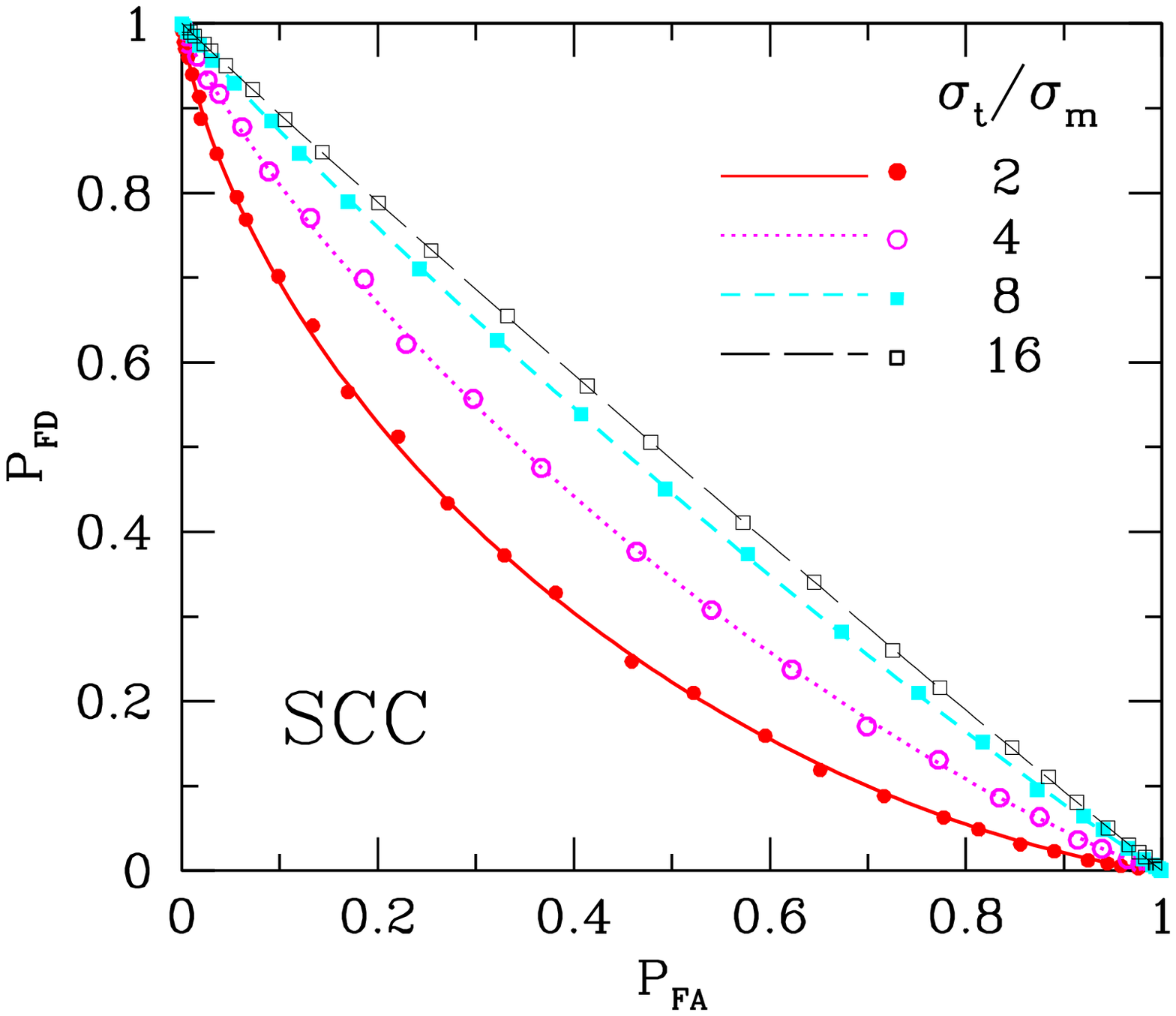}
\end{center}
 \caption{
Same as in Fig.~\ref{fig:pfa-pfd_curve_vary_P}, but we here plot the 
dependence on the ratio $\sigma_{\rm t}/\sigma_{\rm m}$, fixing the 
tail fraction and the amplitude of stochastic signals to $P=0.1$ and 
$\epsilon=0.1$: 
$\sigma_{\rm t}/\sigma_{\rm m}=2$ ({\it filled circles} and {\it solid}); 
$\sigma_{\rm t}/\sigma_{\rm m}=4$ ({\it open circles} and {\it dotted}); 
$\sigma_{\rm t}/\sigma_{\rm m}=8$ ({\it filled squares} and 
{\it short-dashed}); 
$\sigma_{\rm t}/\sigma_{\rm m}=16$ ({\it open squares} and 
{\it long-dashed}). 
 }
     \label{fig:pfa-pfd_curve_vary_sigma}
\end{figure}

Let us first show $P_{\rm FA}$-$P_{\rm FD}$ curves.  
In Fig.~\ref{fig:pfa-pfd_curve_vary_P}, 
the symbols denote 
the simulated $P_{\rm FA}$-$P_{\rm FD}$ curves for GCC ({\it left}) and 
SCC ({\it right}) statistics in a variety of the tail fractions $P$. 
Here, the signal amplitude $\epsilon = 0.1$ and the ratio of the root of
noise variance $\sigma_{\rm t}/\sigma_{\rm m}=4$ are kept fixed. 
Basically, the false-dismissal probability for a given $P_{\rm FA}$ 
becomes large as the tail fraction increases. 
However, for fixed $P$, the false-dismissal probabilities of the 
GCC statistic are always smaller than that of the SCC statistic. 
In left panel of Fig.~\ref{fig:pfa-pfd_curve_vary_P},
the three thick lines indicate the analytic $P_{\rm FA}$-$P_{\rm FD}$
curves without the higher-order terms in Eq.(\ref{gcc1mean}),
which quantitatively agree with the Monte Carlo simulations.  
A closer look at the results for GCC statistic for the tail fraction
$P=0.2$
shows a small discrepancy
between analytic and simulation results, which is mainly attributed to
the higher-order terms neglected in the analytic results.
The thin line in left panel of Fig.~\ref{fig:pfa-pfd_curve_vary_P}
show the same analytic $P_{\rm FA}$-$P_{\rm FD}$
curves, but we take into account
the higher-order terms (\ref{higher}),
where the agreement becomes excellent.
Note that, most of the gravitational-wave detectors have 
a fairly small non-Gaussian component 
and the analytic formulas for $P\ll1$ without the higher-order terms
 would be applicable in practice.

Fig.~\ref{fig:pfa-pfd_curve_vary_sigma} shows another plot of 
the $P_{\rm FA}$-$P_{\rm FD}$ curves. In each panel, fixing 
the tail fraction $P$ to $0.1$, the dependence on the ratio 
$\sigma_{\rm t}/\sigma_{\rm m}$ is depicted, in which  
both the analytic and the simulation results yield the similar trends. 
From this figure, performance of the GCC statistic seems remarkably 
good. Even for larger non-Gaussian tails, 
the $P_{\rm FA}$-$P_{\rm FD}$ curves for GCC statistic almost remain 
unchanged. On the other hand, the SCC statistic gets worse significantly 
as increasing the ratio $\sigma_{\rm t}/\sigma_{\rm m}>1$. 
This is indeed anticipated from the behavior of the quantity 
$\rho_{\rm eff}/\rho$ in Eq. (\ref{eq:ana_pfa-pfd_curve}) 
(see Fig.~\ref{fig:rho_ratio}).

\begin{figure}[tb]
\begin{center}
\includegraphics[width=8cm,angle=0,clip]{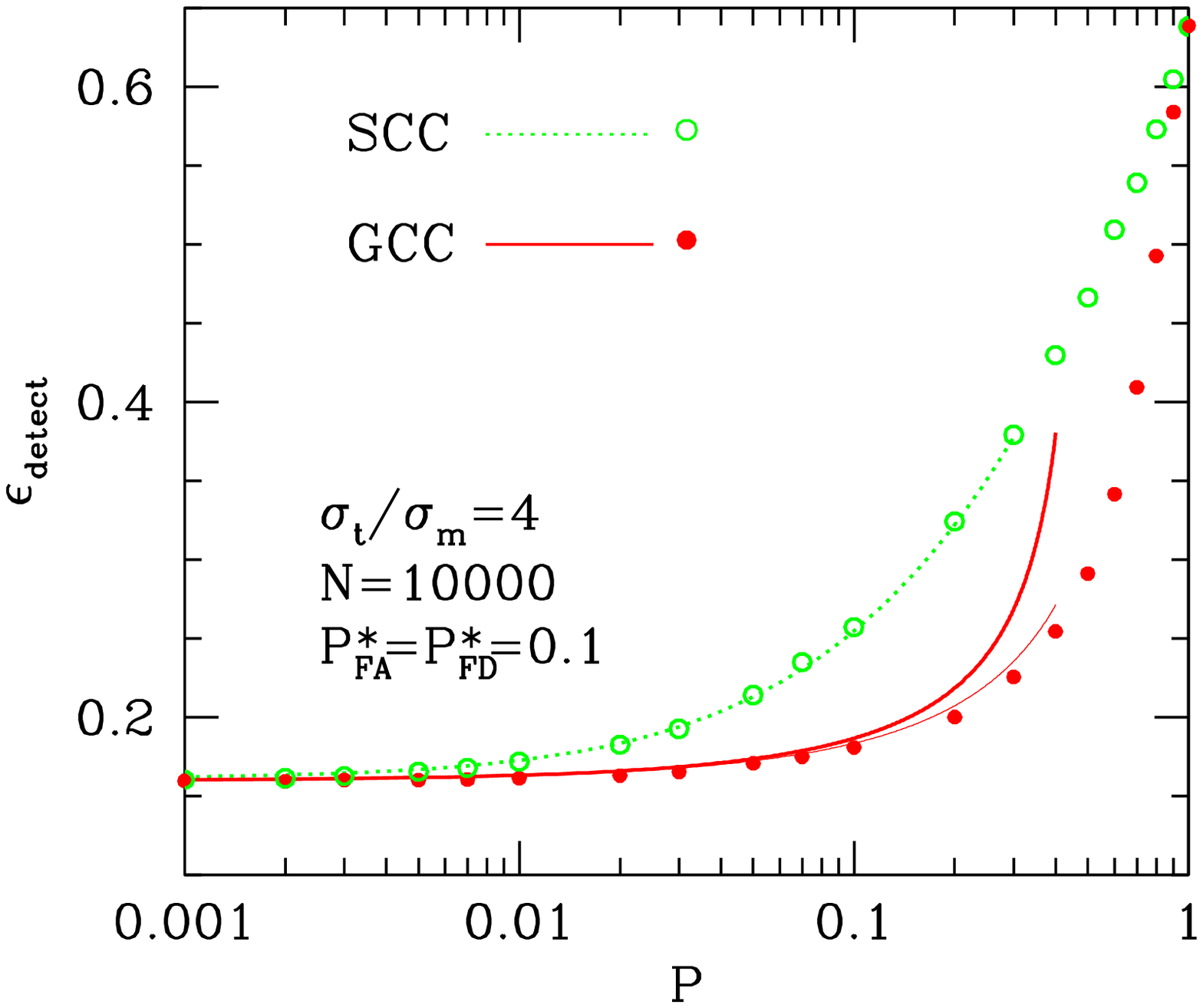}
\hspace*{0.5cm}
\includegraphics[width=8cm,angle=0,clip]{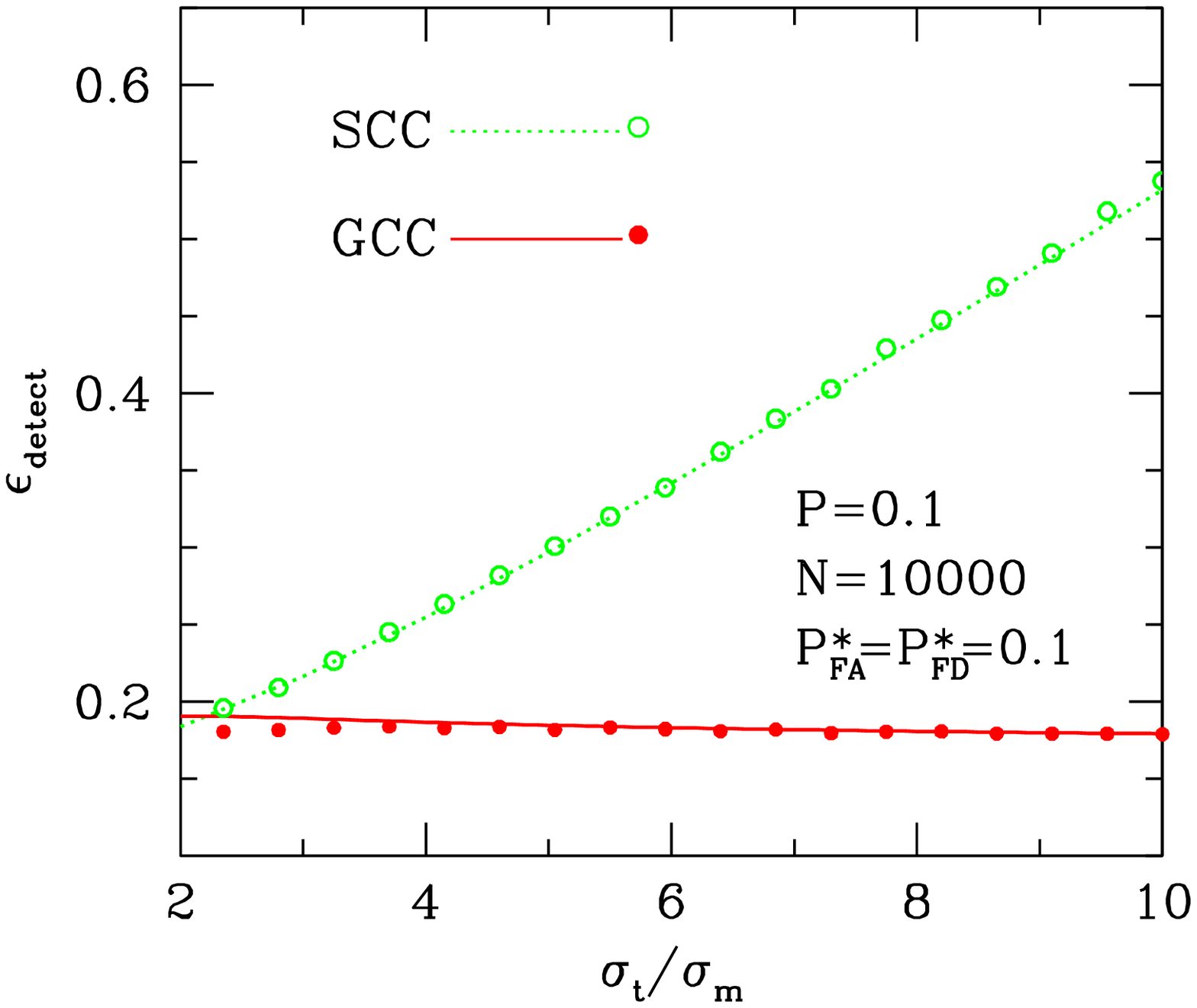}
\end{center}
    \caption{Minimum detectable amplitude of the gravitational-wave signals 
as function of the tail fraction $P$ ({\it left}) and the ratio of noise 
variance $\sigma_{\rm t}/\sigma_{\rm m}$ ({\it right}). 
The ratio of noise variance in left panel is specifically chosen as 
$\sigma_{\rm t}/\sigma_{\rm m}=4$, while the tail fraction in right 
panel is set to $P=0.1$. 
In both panels, filled (open) circles represent the simulation results 
derived from the GCC (SCC) statistic. The corresponding analytic curves 
are also shown in solid and dotted lines based on 
the expressions (\ref{minimum_epsilon}) and (\ref{minimum_epsilon_SCC}).  
The thin line in left panel is the analytical prediction
including the higher-order terms (\ref{higher}).
Note that in these plots, detection point is specifically set to 
$(P_{\rm FA}^{\ast},\,P_{\rm FD}^{\ast})=(0.1,\,0.1)$ with sample points 
$N=10^4$. }
     \label{fig:minimum_detectable_amp}
\end{figure}

Turning to focus on the minimum detectable amplitude, 
we plot in Fig.~\ref{fig:minimum_detectable_amp} the dependence of   
the amplitude $\epsilon_{\rm detect}$ on the tail fraction $P$ 
({\it left}) and the ratio of variance 
$\sigma_{\rm t}/\sigma_{\rm m}$ ({\it right}). In this plot, we specifically 
set the detection point to 
$(P_{\rm FA}^{\ast},\,\,P_{\rm FD}^{\ast})=(0.1,\,\,0.1)$. 
Note that for numerical investigation of the amplitude 
$\epsilon_{\rm detect}$, 
we ran the Monte Carlo simulation 
several times and vary the amplitude $\epsilon$ 
to find the point satisfying the condition 
$(P_{\rm FA},\,\,P_{\rm FD})=(0.1,\,\,0.1)$ until the accuracy with a few percentage has been achieved. 
In each panel, the solid and dotted lines represent the analytic estimates of 
the minimum amplitude for GCC and SCC statistics, respectively 
(Eqs.~(\ref{minimum_epsilon}), (\ref{minimum_epsilon_SCC})).  
The thin line in left panel shows the analytical prediction
including the higher-order terms (\ref{higher}).
For the smaller tail fraction $P\lesssim0.1$, the analytic results 
for GCC statistic reasonably approximate the simulation results and  
the resultant amplitude $\epsilon_{\rm detect}$
is insensitive to the 
non-Gaussian tails. On the other hand, the minimum amplitude of SCC 
statistic increases in linearly proportional to the ratio of noise 
variance $\sigma_{\rm t}/\sigma_{\rm m}$. 
This remarkable feature is precisely what we expected from the analytic 
estimate of the minimum detectable amplitude 
(see Sec.~\ref{subsec:epsilon_min} and Fig.~\ref{fig:gain_factor}). 
That is, the dependence of the ratio $\sigma_{\rm t}/\sigma_{\rm m}$ 
on the functions $G$ and $\epsilon_{\rm detect}^{\rm SCC}$ almost cancels out 
each other, leading to the insensitivity of $\epsilon_{\rm detect}^{\rm GCC}$. 
Since the two-step approximation in our analytic formulas 
becomes a good description for a larger value 
$\sigma_{\rm t}/\sigma_{\rm m}$, 
 as long as the tail fraction $P$ is small,
the analytic estimation of  
$\epsilon_{\rm detect}^{\rm GCC}$ provides a robust and a quantitative 
prediction for the detection efficiency of the GCC.

\section{Summary}
\label{sec:5}

In this paper, we discussed the robust data analysis method to 
detect a stochastic background of gravitational wave in the presence of 
the non-Gaussian noise. Specifically, we have discussed
the generalized cross-correlation (GCC) statistic 
which is a nearly optimal statistic and quantified  
the detection efficiency in an analytic manner. To do this, we have 
focused on a simple but realistic non-Gaussian noise 
model, i.e., two-component Gaussian noise. 
We derived the analytic formulas for the false-alarm and the 
false-dismissal probabilities as a function of threshold value 
$\Lambda_{\ast}$ and obtained the $P_{\rm FA}$-$P_{\rm FD}$ curves. 
Also, we derived the minimum detectable amplitude of stochastic 
signal, $\epsilon_{\rm detect}$. 
These analytic results are compared with the Monte Carlo simulations for 
the cross-correlation analysis and found that the analytic formulas 
provide a good description.

For small tail fraction $P_i\lesssim0.1$,  
from Eqs.~(\ref{minimum_epsilon})--(\ref{gain}), 
minimum detectable amplitude of the 
stochastic signal for GCC statistic is related to that of the SCC statistic: 
\begin{eqnarray}
\epsilon_{\rm detect}^{\rm GCC}
\simeq 
\left[\{1+P_1\,(\sigma_{{\rm t},1}/\sigma_{{\rm m},1})^2\}
\{1+P_2\,(\sigma_{{\rm t},2}/\sigma_{{\rm m},2})^2\}\right]^{-1/4} \,\,
\epsilon_{\rm detect}^{\rm SCC}
\nonumber
\end{eqnarray}
where the quantity $\epsilon_{\rm detect}^{\rm SCC}$ become
\begin{eqnarray}
\epsilon_{\rm detect}^{\rm SCC} \simeq 
\left\{\frac{2\sqrt{2}\,\,\gamma}{\sqrt{N}}\,\,
\sigma_{{\rm m},1}\sigma_{{\rm m},2}\right\}^{1/2}\,\,
\left[\{1+P_1\,(\sigma_{{\rm t},1}/\sigma_{{\rm m},1})^2\}
\{1+P_2\,(\sigma_{{\rm t},2}/\sigma_{{\rm m},2})^2\}\right]^{1/4} 
\nonumber
\end{eqnarray}
with $\gamma$ being $\gamma=\rm{erfc}^{-1}[2P_{\rm FA}]$. 
Thus, these two equations indicate that the minimum amplitude of 
GCC statistic is mainly determined by the main part and is 
insensitive to the tail part of the noise probability distribution. 
Therefore, the quantity $\epsilon_{\rm detect}^{\rm GCC}$ is  
almost equivalent to the one derived from the SCC statistic just dropping 
the contribution of non-Gaussian tails: 
\begin{eqnarray}
\epsilon_{\rm detect}^{\rm GCC} \simeq 
\left\{\frac{2\sqrt{2}\,\,\gamma}{\sqrt{N}}\,\,
\sigma_{{\rm m},1}\sigma_{{\rm m},2}\right\}^{1/2}. 
\nonumber
\end{eqnarray}

Finally, we close this paper with comments and discussions.
Throughout the paper, we have considered the two coincident 
and coaligned detectors with the white noise spectra. 
In practice, these restrictions must be relaxed.
According to 
Refs.~\cite{Allen:2001ay,Allen:2002jw},  
the GCC statistic has been extended to deal with a more 
realistic situation with non-coincident and non-co-aligned detectors of 
the colored noises. In this context, 
the analysis in the present paper roughly matches the 
narrow-band analysis 
in the Fourier domain, where the noise spectrum can be approximately 
described by a white noise. 
The extension of the present analysis to the broad-band case would be
straightforward and this should deserve consideration.
Another important simplification in our analysis is the stationarity
of the instrumental noises and neglect of a noise correlation 
between two detectors. 
In practice, the noise correlation is known 
as a big obstacle in the LIGO at the Hanford site \cite{Lazzarini:2004hk} 
and it would potentially be a serious problem in the future detector, 
LCGT \cite{Mio:2003ii}. 
Thus, exploration of optimal data analysis strategy in the presence of
not only the non-Gaussian noise but also the nonsteady noise and the
noise correlation is very important task for future detectors.

It will be rather difficult to improve the sensitivity of the detectable 
amplitude by building a more sophisticated detector, due to the limitation of available technology and funds.  
Hence, efficient methods for data analysis such as the GCC statistic 
should be further exploited and it must be 
properly incorporated into the future detection of stochastic 
gravitational waves. 
Extending the present work to deal with a more realistic situation, 
we will continue to address these issues.

\acknowledgments
We are grateful to the anonymous referee for 
his constructive suggestions and many useful comments, which improve the 
original manuscript, especially in the  
analytic treatment of the detection efficiency (Sec. 
\ref{subsec: mean_var_GCC},\ref{subsec:epsilon_min} and 
Appendix \ref{appen:appendix1}). 
We would like to thank Takahiro Tanaka for useful discussions. 
Y.H. thanks Hirotaka Takahashi, Koji Ishidoshiro and Wade Naylor 
for helpful comments. 
Y.H. also thanks Bruce Allen for fruitful discussions in Amaldi 6.
Y.H., H.K. and T.H. were supported by 
a Japan Society for the Promotion of Science (JSPS) Research Fellowships. 
A.T. acknowledges the support by a Grant-in-Aid for Scientific 
Research from the JSPS (No.~18740132). 

\appendix
\section{Analytical expressions for the means 
  and the variances for the GCC statistic}
\label{appen:appendix1}

In this Appendix, we derive the analytical expressions 
(\ref{gcc0meantrue})--(\ref{gcc1variancetrue})
for the mean and the variance of the GCC statistic.

First, we compute the mean and the variance 
in the absence of signal, i.e., ${\cal T}=0$.
Adopting the two-step approximation (\ref{twostep}) with the critical 
value (\ref{critical}), we obtain
\begin{eqnarray}
\langle \Lambda_{\rm GCC}^{(0)} \rangle &=& 0, \label{gcc0mean}\\
\left[\Delta\Lambda_{\rm GCC}^{(0)} \right]^{2}
&=& \frac{\langle n_{1}^{2} \rangle_{\rm G} \langle n_{2}^{2}\rangle_{\rm G} }
{{N}}, 
\label{gcc0variance}
\end{eqnarray}
where
\begin{eqnarray}
\langle n_{i}^{2} \rangle_{\rm G} 
&=& \int_{-x_{{\rm cr},i}}^{x_{{\rm cr},i}}
     dn_{i}\, \Bigl[   n_{i}^{2} \,p_{n,i}(n_{i}) \Bigr]
 + 2 \left(\frac{\sigma_{{\rm m},i}}{\sigma_{{\rm t},i}}\right)^{4}
\int_{x_{{\rm cr},i}}^{\infty} dn_{i} \Bigl[   n_{i}^{2} \,p_{n,i}(n_{i}) \Bigr].
\label{noise_mean_1}
\end{eqnarray}
In the situation we are interested in, i.e., $P_{i} \ll 1$ and 
$(\sigma_{{\rm m},i}/\sigma_{{\rm t},i}) \lesssim 1$, 
the contribution of second term 
in the right hand side of Eq.(\ref{noise_mean_1}) is negligibly small.
Thus, the variance of noise is approximately described by the first term.
In Ref.~\cite{Allen:2001ay}, this effect has been called {\it clipping}. 
Then, we have
\begin{eqnarray}
\langle n_{i}^{2} \rangle_{\rm G} &\approx& \int_{-x_{{\rm cr},i}}^{x_{{\rm cr},i}}
 dn_{i}
 \Bigl[   n_{i}^{2} \,p_{n,i}(n_{i}) \Bigr]
  \notag \\
&=&
(1-P_{i}) \, \sigma_{{\rm m},i}^{2} \,
P_{{\rm G}}[x_{{\rm cr},i},\sigma_{{\rm m},i}]
+ P_{i} \, \sigma_{{\rm t},i}^{2} \,
P_{{\rm G}}[x_{{\rm cr},i},\sigma_{{\rm t},i}] \,.
\label{noise_mean_app}
\end{eqnarray}
Here, the quantity $P_{{\rm G}}[x,\sigma]$ is defined in 
Eq.(\ref{define1}):
\begin{eqnarray}
&&P_{{\rm G}}[x,\sigma] \equiv  
   {\rm erf}
\left[\frac{x}{\sqrt{2}\sigma} \right]  
-\sqrt{\frac{2}{\pi}}
\frac{x}{\sigma} e^{
-(x/\sigma)^{2}/2} \,.  \notag 
\end{eqnarray}

Next, we consider the mean and the variance in the presence of 
gravitational-wave signals. The mean 
$\langle \Lambda_{\rm GCC}^{(1)} \rangle $
is expressed as 
\begin{eqnarray}
\langle \Lambda_{\rm GCC}^{(1)} \rangle 
&=& \frac{1}{N} \sum_{k=1}^{N} \int  ds_{1}^{k}\,ds_{2}^{k} \,
f_{1}'(s_{1}^{k}) \cdot f_{2}'(s_{2}^{k}) \, p_{s}(s_{1}^{k},s_{2}^{k})\, ,
\label{eq:mean_GCC_1}
\end{eqnarray}
where $p_{s}(s_{1}^{k},s_{2}^{k})$ is 
the joint probability distribution function
 for the two detector outputs defined by 
\begin{equation}
 p_{s}(s_{1}^{k},s_{2}^{k}) \equiv 
\int  dh^{k}\,dn_{1}^{k}\,dn_{2}^{k} \, 
\delta(s_{1}^{k}-h^{k}-n_{1}^{k}) \, \delta(s_{2}^{k}-h^{k}-n_{2}^{k})
\,p_{h^{k}}(h^{k})\,p_{n,1}(n_1^{k})\,p_{n,2}(n_2^{k}).
\end{equation}
As long as the two-step approximation with clipping holds, 
the quantity (\ref{eq:mean_GCC_1}) up to 
$\mathcal{O}(\epsilon^{2})$
becomes
\begin{equation}
\langle \Lambda_{\rm GCC}^{(1)} \rangle  \approx 
\langle \Lambda_{\rm GCC}^{(1)} \rangle^{({\rm \ell})} +
\langle \Lambda_{\rm GCC}^{(1)} \rangle^{({\rm h})}\,,
\label{gcc1mean_appendix}
\end{equation}
where,
\begin{eqnarray}
 \langle \Lambda_{\rm GCC}^{(1)} \rangle^{({\rm \ell})} =
\epsilon^{2}\{ 1-(P_{1}+P_{2}) \}
P_{{\rm G}}[x_{{\rm cr},1},\sigma_{{\rm m},1}]\,
P_{{\rm G}}[x_{{\rm cr},2},\sigma_{{\rm m},2}]
\end{eqnarray}
and
\begin{eqnarray}
\langle \Lambda_{\rm GCC}^{(1)} \rangle^{({\rm h})} 
&=&
\epsilon^{2}(
P_{1}\,P_{{\rm G}}[x_{{\rm cr},1},\sigma_{{\rm t},1}]\,
P_{{\rm G}}[x_{{\rm cr},2},\sigma_{{\rm m},2}]+
P_{2}\,P_{{\rm G}}[x_{{\rm cr},1},\sigma_{{\rm m},1}]\,
P_{{\rm G}}[x_{{\rm cr},2},\sigma_{{\rm t},2}])\, \notag \\
&+& \epsilon^{2}P_{1}\,P_{2}\,(P_{{\rm G}}[x_{{\rm cr},1},\sigma_{{\rm m},1}]\,
P_{{\rm G}}[x_{{\rm cr},2},\sigma_{{\rm m},2}]-
P_{{\rm G}}[x_{{\rm cr},1},\sigma_{{\rm t},1}]\,P_{{\rm G}}[x_{{\rm cr},2},\sigma_{{\rm m},2}] \notag \\
&-&P_{{\rm G}}[x_{{\rm cr},1},\sigma_{{\rm m},1}]\,P_{{\rm G}}[x_{{\rm cr},2},\sigma_{{\rm t},2}]
+P_{{\rm G}}[x_{{\rm cr},1},\sigma_{{\rm t},1}]\,P_{{\rm G}}[x_{{\rm cr},2},\sigma_{{\rm t},2}])\,.
\label{higher}
\end{eqnarray}
Under the situation that 
$P_{i} \ll 1$ and 
$(\sigma_{{\rm m},i}/\sigma_{{\rm t},i}) \lesssim 1$,
the critical value $x_{{\rm cr},i}$ defined in Eq.(\ref{critical})
satisfies the condition 
$\sigma_{{\rm m},i} \ll  x_{{\rm cr},i} \ll \sigma_{{\rm t},i}$,
then $P_{{\rm G}}[x_{{\rm cr},i},\sigma_{{\rm m},i}]$ and
$P_{{\rm G}}[x_{{\rm cr},i},\sigma_{{\rm t},i}]$ approximately become
unity and zero, respectively.
Therefore,
one can regard the term $\langle \Lambda_{\rm GCC}^{(1)} \rangle^{({\rm h})}$ 
as the negligible higher-order terms. 

Finally, using the two-step approximation with clipping, 
the leading order result of the quantity 
$\Delta\Lambda_{\rm GCC}^{(1)}$ becomes 
\begin{eqnarray}
\Delta\Lambda_{\rm GCC}^{(1)} &=&
\sqrt{\langle(\Lambda_{\rm GCC}^{(1)})^{2}\rangle-
\langle\Lambda_{\rm GCC}^{(1)}\rangle^{2}} \\
 &\approx & 
\frac{(\langle n_{1}^{2} \rangle_{\rm G}\langle n_{2}^{2} \rangle_{\rm G})^{1/2}}
{\sqrt{N}} \left(1+\mathcal{O}
\left(\frac{\epsilon^{2}}{\langle n_{i}^{2} \rangle_{\rm G}}\right)\right).
%
\label{gcc1variance}
\end{eqnarray}
Thus, in the present situation that the detector noises dominate
the gravitational signal, we can reasonably treat the quantity 
$\Delta\Lambda_{\rm GCC}^{(1)}$ as
\begin{equation}
 \Delta\Lambda_{\rm GCC}^{(1)} \approx 
\frac{(\langle n_{1}^{2} \rangle_{\rm G}\langle n_{2}^{2} \rangle_{\rm G})^{1/2}}
{\sqrt{N}} = \Delta\Lambda_{\rm GCC}^{(0)} \,.
\end{equation}
%




\end{document}